\documentclass[%
 reprint,
 aip,
 graphicx,
 amsmath,amssymb,
]{revtex4-2}

\usepackage{graphicx}
\usepackage{dcolumn}
\usepackage{bm}

\usepackage{color}
\usepackage[flushleft]{threeparttable}

\usepackage[hidelinks]{hyperref}
\usepackage[capitalise]{cleveref}
\usepackage[utf8]{inputenc}
\usepackage[T1]{fontenc}
\usepackage{mathptmx}
\usepackage{etoolbox}
\makeatletter
\def\@email#1#2{%
 \endgroup
 \patchcmd{\titleblock@produce}
  {\frontmatter@RRAPformat}
  {\frontmatter@RRAPformat{\produce@RRAP{*#1\href{mailto:#2}{#2}}}\frontmatter@RRAPformat}
  {}{}
}%
\makeatother

\newcommand{\citeinline}[1]{Ref. \onlinecite{#1}}
 
\begin{document}

\preprint{AIP/123-QED}

\title[Controlling the motional quality factor of a diamagnetically levitated graphite plate]{Controlling the motional quality factor of a diamagnetically levitated graphite plate}
\author{P. Romagnoli}
\thanks{These three authors contributed equally}
\affiliation{ 
Quantum Machines Unit, Okinawa Institute of Science and Technology Graduate University, Onna, Okinawa 904-0495, Japan \looseness=-1
}%
\author{R. Lecamwasam}
\thanks{These three authors contributed equally}
\affiliation{ 
Quantum Machines Unit, Okinawa Institute of Science and Technology Graduate University, Onna, Okinawa 904-0495, Japan \looseness=-1
}
\author{S. Tian}
\thanks{These three authors contributed equally}
\affiliation{ 
Quantum Machines Unit, Okinawa Institute of Science and Technology Graduate University, Onna, Okinawa 904-0495, Japan \looseness=-1
}%
\author{J.E. Downes}
\affiliation{
School of Mathematical and Physical Science, Macquarie University, 2109 NSW, Australia \looseness=-1
}

\author{J. Twamley}%
\email{jason.twamley@oist.jp}
\affiliation{ 
Quantum Machines Unit, Okinawa Institute of Science and Technology Graduate University, Onna, Okinawa 904-0495, Japan \looseness=-1
}%

%
%

\date{\today}

\begin{abstract} \label{abs}
Researchers seek methods to levitate matter for a wide variety of purposes, ranging from exploring fundamental problems in science, through to developing new sensors and mechanical actuators. Many  levitation techniques require active driving and most can only be applied to objects smaller than a few micrometers. Diamagnetic levitation has the strong advantage of being the only form of levitation which is passive, requiring no energy input, while also supporting massive objects. Known diamagnetic materials which are electrical insulators are only weakly diamagnetic, and require large magnetic field gradients to levitate. Strong diamagnetic materials which are electrical conductors, such as graphite, exhibit eddy damping, restricting motional freedom and reducing their potential for sensing applications. In this work we describe a method to engineer the eddy damping while retaining the force characteristics provided by the diamagnetic  material. We study, both experimentally and theoretically, the motional damping of a magnetically levitated graphite plate in high vacuum and demonstrate that one can control the eddy damping by patterning the plate with through-slots which interrupt the eddy currents. We find we can control the motional quality factor over a wide range with excellent agreement between the experiment and numerical simulations.


\end{abstract}

\maketitle




The levitation of matter has a variety of applications ranging from quantum science and technology through to industrial development of levitated actuators, motors, and robots. Typically to levitate or trap an object requires a source of power, e.g. optical trapping uses strong laser fields, while dynamic electric fields can hold charged objects in Paul or Penning traps. These so-called ``active'' systems suffer from various types of noise, and thereby motional heating. In contrast ``passive'' diamagnetic levitation, which requires no active driving or energy input, has the potential for very low noise, low heating, levitation and motional control. Moreover, macroscopic objects can easily be supported and manipulated \cite{Chen2020}. For this reason diamagnetic levitation has recently attracted much attention and since its original exposition \cite{Berry1997,Simon2000, Simon2001}, it is now firmly regarded as one of the main techniques in {\em levitodynamics} \cite{Gonzalez-Ballestero2021, Gao2021MagneticProspects,Millen2020QuantumParticles}. 

Levitation can be used to explore many fundamental questions, including testing ideas in thermodynamics \cite{Gieseler2018} and alternative theories of quantum mechanics \cite{Pontin2020}, and searching for new types of forces in nature \cite{Moore2021, Xiong2021SearchingSensor, Timberlake2021ProbingResonators}. Tabletop experiments have been proposed which explore the relationship between quantum theory and gravity \cite{Bose2017,Marletto2017, Christodoulou2019, Carlesso2019, Nguyen2020, Marshman2022ConstructingInterferometer}. Another major application of levitodynamics is the development of novel sensors, including inertial sensors \cite{Garmire2007, Hempston2017, Prat-camps2017, Timberlake2019, Monteiro2020, Lewandowski2021},
detecting gravitational waves \cite{Arvanitaki2013DetectingSensors} and dark matter/energy \cite{Yin2022ExperimentsEnergyc}, magnetometry \cite{Kumar2017},  measuring mass \cite{Chen2021DiamagneticallyScaleb}, and light pressure sensing \cite{Vaskuri2021AbsoluteMass}. It is only in levitated systems  (using optical trapping) that a macroscopic quantum superposition has ever been achieved \cite{Tebbenjohanns2020MotionalSpace, Delic2020CoolingState, Magrini2021Real-timeTemperature, Tebbenjohanns2021QuantumSpace}. Within levitodynamics, graphite plates are gaining increasing attention as a highly promising platform. These have already been used to experimentally test theories of dark matter \cite{Yin2022ExperimentsEnergyc}, and may outperform other types of levitated systems \cite{Chen2022DiamagneticResonators}.

Besides quantum science, another important area of levitodynamics is the development of levitated actuators, motors, and robots. This mostly uses magnetic levitation, due to its ability to support large mass loads against gravity. An example includes the development of  miniature robots which move on a planar surface and can be controlled via localised currents \cite{Pelrine2012DiamagneticallyProperties, Pelrine2016OptimalPatterns,Hsu2017AutomatedMilli-robots,Pelrine2019MethodsDesigns, Kuthan2020CollectiveOperation}. Researchers have recently developed photo-thermal methods to actuate the motion of a diamagnetically levitated graphite plate, which has led to a large interest in developing photo-activated 2D positioners  \cite{Kobayashi2012OpticalGraphite,Han2018Carbon-BasedActuators,Yang2019PhotoresponsiveMaterials,Young2019OpticalGraphite,Yang2019PhotoresponsiveMaterials,Ewall-Wice2019OptomechanicalGraphite, Kim2020HighlyMultistimuliResponses,
Huang2021Light-responsiveApplications, Yee2021PhotothermalRevised}. We particularly note recent works using diamagnetic levitated graphite for optical energy harvesting \cite{Shen2022OpticalGraphite}, and a multiple-degree-of-freedom nanopositioner \cite{Vikrant2022DiamagneticallyFreedom}. 

Diamagnetic materials which are electrical insulators, including diamond, polymers, some glasses, and many organic materials \cite{Lewandowski2021High-SensitivityMicrosphere,Nakashima2020, Leng2021}, tend to be only very weakly diamagnetic. 
For diamagnetic levitation of macroscopic masses, for example those used in actuators, the ideal material is Highly Oriented Pyrolytic Graphite (HOPG). The magnetic susceptibility of HOPG is highly aniostropic, with the 
direction of strongest diamagnetic susceptibility normal to the slab face \cite{Niu2018, Chen2020}. However, HOPG is an excellent conductor, and thus eddy currents are induced as it moves through magnetic fields. These damp the motion, reducing the quality factor of the motional mode, and producing heat leading to thermal noise.

For applications such as robots and actuators it may not be necessary to completely eliminate eddy damping, which may even be useful in some schemes. For other applications however, such as motional sensors, it is desirable to completely remove the eddy damping. 
In electrical transformers, engineers reduce eddy losses either by interrupting the currents with layers of laminated material, or by making the core of the transformer out of a highly resistive ferrite magnetic material. The latter route was recently adapted to the diamagnetic levitation of graphite \cite{Chen2022DiamagneticResonators}, where micrometer-scale graphite particles were encapsulated in an electrically insulating resin. Eddy currents and the associated damping were significantly reduced, leading to quality factors of $5\times 10^5$. At the same time however, this results in a lowering of the diamagnetic lift. The filling fraction of graphite within the resin cannot be too high without compromising structural integrity, and random orientation of the particles within the resin lowers the effective magnetic susceptibility.

In our work we seek to control eddy damping using a method more similar to the lamination technique in electrical transformers. We consider the diamagnetic levitation of a solid slab of HOPG, which we then pattern with very narrow through-cut slots, the purpose of which is to interrupt the path of eddy currents. We machine several samples with increasing slot densities. We hypothesise that as we modify the density of slotted interruptions to the eddy currents, then these currents will be modified, and so too will the associated eddy damping. The advantage of this method to control the eddy damping is that one retains virtually all the diamagnetic lift of the original HOPG slab, as the through-slots are very narrow. 

In the following 
we outline the experiment, analysis, and simulation of the levitodynamics of the slotted graphite plates. We find that the through-cut slots allow us to systematically control the motional quality factor in a highly predictable manner, with excellent agreement between theory and experiment.

\begin{figure}
    \centering
		\includegraphics[width=\linewidth]{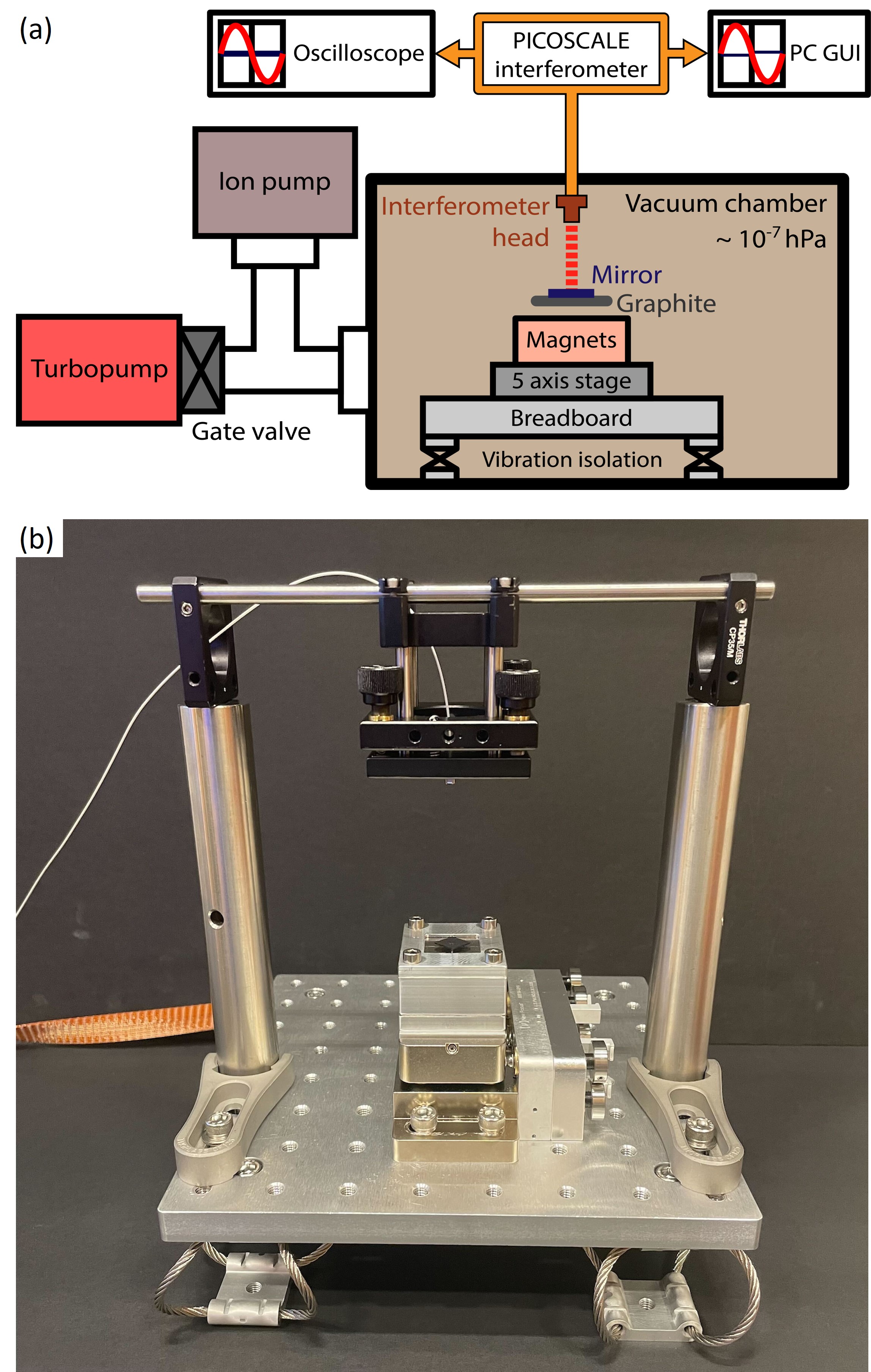}
\caption{Experimental setup. (a) A plate of pyrolitic graphite is levitated by four NdFeB N52 magnets with polarities arranged in a checkerboard pattern. The magnets are fixed on a five-axis motorized stage, which is supported by a breadboard resting upon four vibration isolation mounts within the vacuum chamber. The vacuum chamber and ion pump lie on a vibration isolation optical table, while the turbopump is on a separate vibration isolation platform. A small mirror is fixed to the graphite sample, which is used by an interferometer to measure the vertical displacement. The interferometer is aligned using the five-axis stage. (b) Photograph of the platform which sits inside the vacuum chamber.} \label{fig:setup}
\end{figure}
\begin{figure*}
   \includegraphics[width=\linewidth]{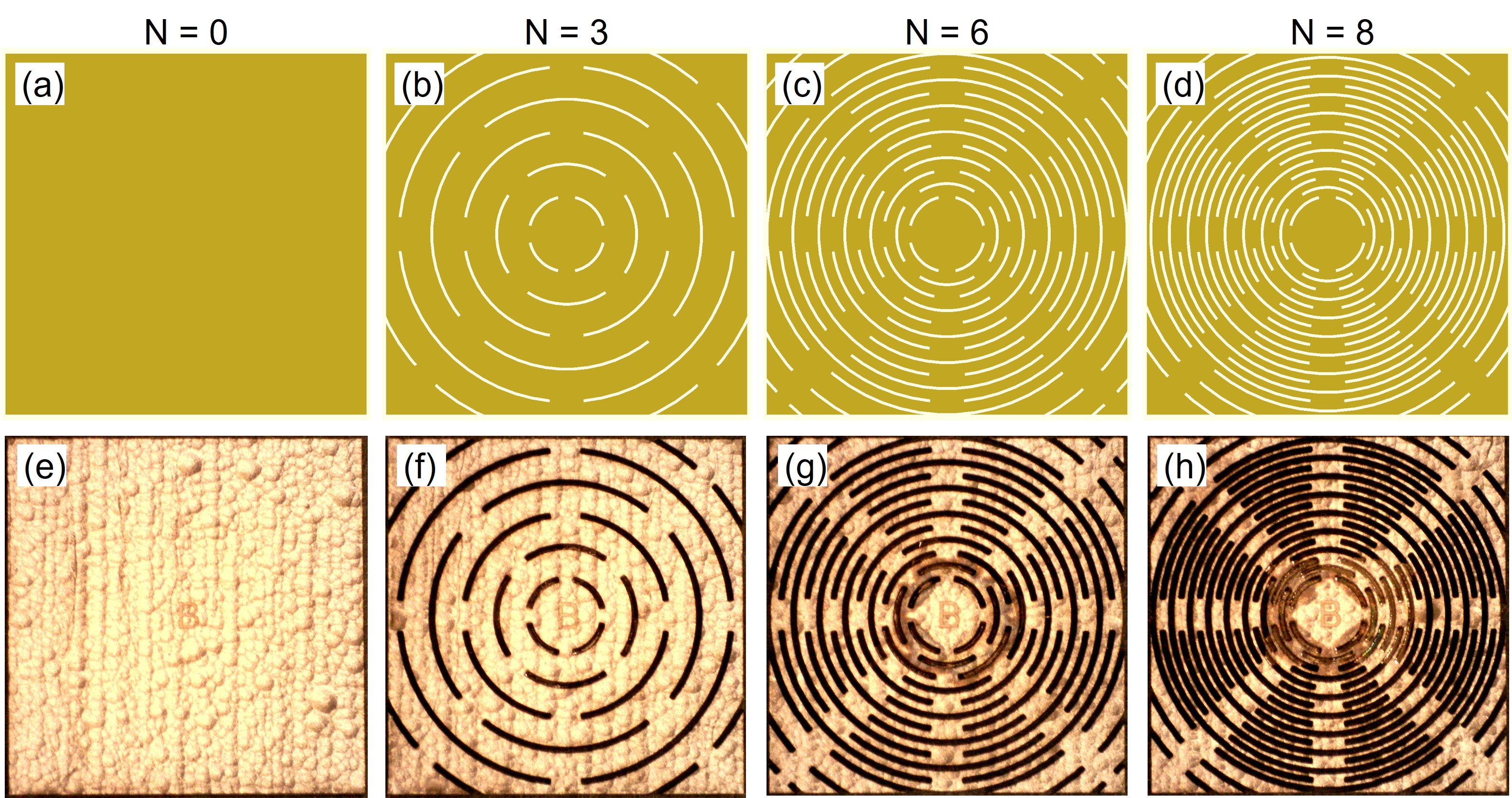}
    \caption{Slit designs and photographs of the machined graphite samples. (a)-(d) CAD design patterns of ring-like slits. These are named $N=0$, $3$, $6$, and $8$, with $N$ reflecting a parameter used to generate the ring patterns. As $N$ grows larger slit density increases, leading to stronger suppression of eddy currents and an increase in motional quality factor.   (e)-(h) Photos of the machined graphite plates, measuring $10\,\mathrm{mm}\times 10\,\mathrm{mm}$. The thickness is approximately $0.77\,\mathrm{mm}$, however this is not uniform due the visible surface roughness. After machining, a mirror (not shown, made from aluminum coating on glass) was glued to the center of each plate for optical measurement of the motion. } \label{fig:SlitPattern} 
\end{figure*}
A sketch of the experimental setup is presented in \cref{fig:setup}.
A plate of pyrolytic graphite is levitated by four NdFeB magnets  placed in an alternating polarity checkerboard pattern.
The magnets are rigidly held within a holder which is fixed to a five axis vacuum-compatible motorised stage. The magnet/motor platform is mounted on a small optical breadboard which in turn sits on four vibration isolation supports.
The position of the graphite sample is monitored by an interferometric displacement sensor.
This displacement sensor is based on a compact Michelson interferometer, and enables high precision measurements in real time with a resolution of picometers at a high bandwidth.

The whole structure is positioned in a vacuum chamber, which is evacuated by a system consisting of a turbopump, an ion pump, and associated roughing pump.
During the measurement periods the turbopump is switched off to avoid unwanted mechanical vibrations, while the ion-pump operates continuously to maintain high vacuum ($10^{-7}\:{\rm hPa}$). 
The vacuum chamber and ion pump are supported by a damped and vibration isolated optical table, while the turbopump is supported by a separate vibration damped and isolated platform.

The experiment aims to increase the motional quality factor of a diamagnetically levitated slab of pyrolytic graphite. Four $10\,{\rm mm}\times 10\,{\rm mm}$ samples were machined from a single piece of graphite, to ensure they all possessed similar electric and magnetic properties. The thickness of each sample was approximately $0.77\,{\rm mm}$, however this was not uniform since the graphite surface was very coarse. Each plate had a pattern of ring-like slits machined into it whose purpose was to interrupt the eddy currents, and hence lower the resulting eddy damping forces. These slits were created by femtosecond laser machining. The slit designs, and photographs of the machined samples, are shown in \cref{fig:SlitPattern}.
A small mirror was then glued onto the centre of each piece, to allow an interferometer to measure vertical displacement. 

Each graphite sample was levitated for a period of twenty minutes, and its vertical position recorded using the interferometer. The resulting power spectral density was then analysed as described below. 
Further details of the setup are given in the Supplementary Material \S 1.

\begin{figure}
	\includegraphics[width=\linewidth]{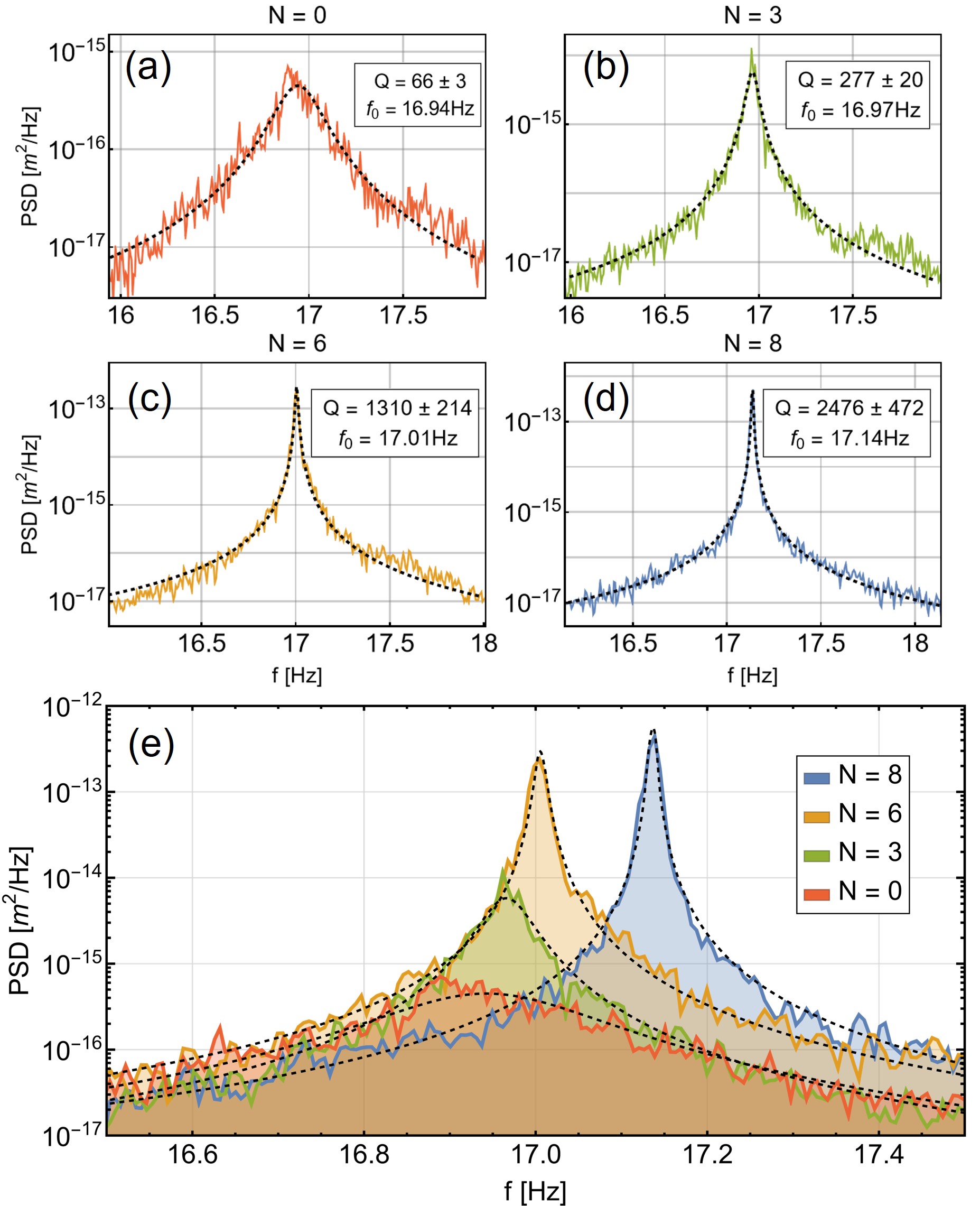}
\caption{Power spectral density of the vertical motion. In (a)-(d) the solid, coloured traces show the experimental data. The dashed black line denotes the theoretical fit, with insets showing the fitted quality factors ($Q$) and resonance frequencies ($f_0$). Errors come from the covariance matrix of the fit. With increased slit density, the quality factor increases from $Q_{N=0}\sim 66$ to $Q_{N=8}\sim 2476$. In (e) we overlap the power spectral densities. As the number of slits increases the resonance peaks get sharper, and there is a slight increase in frequency.} \label{fig:PSD}
\end{figure}
Due to its positioning, the interferometer is most sensitive to vertical motion of the plate. 
Using the normal mode simulation methods outlined above we find,
and in agreement with Chen et al. \cite{Chen2020}, motion corresponding to three motional modes: vertical oscillation along the $z$-axis, and torsion about the $x$- and $y$-axes. These are all expected to have frequencies of around 17Hz.

As the graphite plate moves, the magnetic field it experiences, and hence the magnitude and direction of the force on the plate, vary in a complex manner. This will cause a nontrivial coupling between all of the motional modes. For simplicity, we will approximate the three vertical modes as a single effective mode of a one-dimensional oscillator. This oscillator is primarily damped by eddy currents, induced by motion of the plate through the magnetic field. As we discuss in the Supplementary Material \S 3, damping due to air should be negligible at the pressures we consider.

The power spectral density for a harmonic oscillator is given in  \S10 of \citeinline{Wang1945OnII}, and discussed further in our Supplementary Material \S 3. The power spectral densities for the experimental data traces were fitted to the theoretical values, allowing us to extract the effective damping rates $\gamma$ and natural frequencies $f_0$, from which we could calculate the quality factor $Q$ of each oscillator. These are shown in \cref{fig:PSD}. 

As expected, the oscillation frequencies were all approximately 17Hz. As the number of slots increases, there is a slight upwards shift in $f_0$. This is most likely due to the slots removing material from the graphite plate, leading to a decrease in levitation height, and hence change in the magnetic trapping force. In total, the rings are able to increase the oscillator quality factor by a factor of approximately forty.


To understand the frequencies observed in the power spectral density, we simulated the motional modes for the $n=0$ plate. These modes are determined by the forces and torques experienced as it moves through the inhomogenous magnetic field above the magnet array.  As discussed in the Supplementary Material \S 2, there are six modes in total. Three of them, oscillations in the horizontal plane or rotation about the vertical axis, have frequencies bunched around $4\,\mathrm{Hz}$. The other three involve vertical motion, namely vertical oscillation and tilting about the horizontal axes, and are predicted to have frequencies around $17\,\mathrm{Hz}$. It is these vertical modes that our interferometric setup will be most sensitive to. 

To estimate the increase in quality factor due to the slits, we simulated the eddy currents in each of the graphite plates in \cref{fig:SlitPattern}. The currents are induced by an effective electric potential which depends on the geometry of the plate and its motion through the external magnetic field \cite{Kirpo2010Eddy-currentDipole,Votyakov2012InteractionPlate,Carlstedt2014ComparisonSimulations,Chen2020}. This current then exerts a force on the plate due to the magnetic field, which can be integrated to find the induced eddy damping. We developed both a two-dimensional model in Mathematica, and three-dimensional COMSOL simulation, the details of which are described in Supplementary Material \S 4. 

The two-dimensional model used the finite element method in Mathematica 13.0 to simulate the currents. Moving from $n=0$ to $n=8$, the total current was found to have decreased by an order of magnitude. Around the edges of a slit very large currents could occur, however as these occur in infinitesimal areas they do not contribute significantly to eddy damping. 

\begin{figure}
    \includegraphics[width=\linewidth]{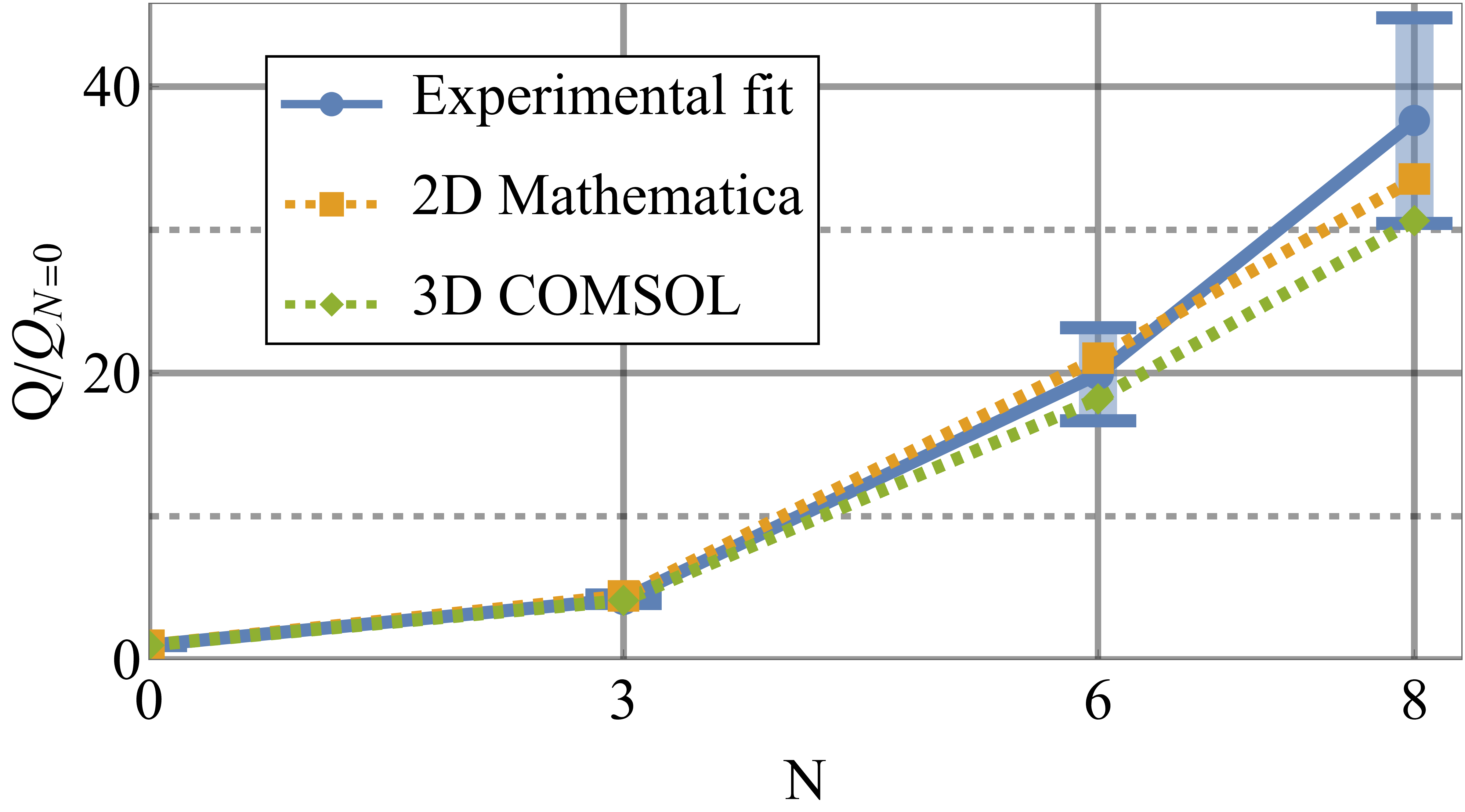}
    \caption{Plate quality factors relative to $n=0$. The solid blue trace is obtained from fitting the experimental power spectral density. The dashed lines show the simulated quality factors due to eddy damping. The orange trace shows the two-dimensional model in Mathematica, while the green is the three-dimensional COMSOL model. The simulations are all within experimental error, indicating that the observed increase in quality factor is due to suppression of eddy currents.}\label{fig:qratios}
\end{figure}

At low pressures, eddy currents are solely responsible for motional damping $\gamma_{\rm eddy}$ of the plate. The quality factor of a mode is inversely proportional to the damping rate: $q_n\propto m_n/\gamma_{\rm eddy\,n}$, for sample $n$ with mass $m_n$. Hence we can predict the ratio of quality factors of the different plates, by calculating the ratios of their corresponding $\gamma_{\rm eddy}$ values and masses. These are shown in \cref{fig:qratios}. We can see that the simulated values agree well with experiment. The $n=8$ sample does appear to have a slightly larger quality factor than predicted, which as we will discuss later is most likely due to the machined slots being wider than the design.

\begin{figure*}
    \includegraphics[width=\linewidth]{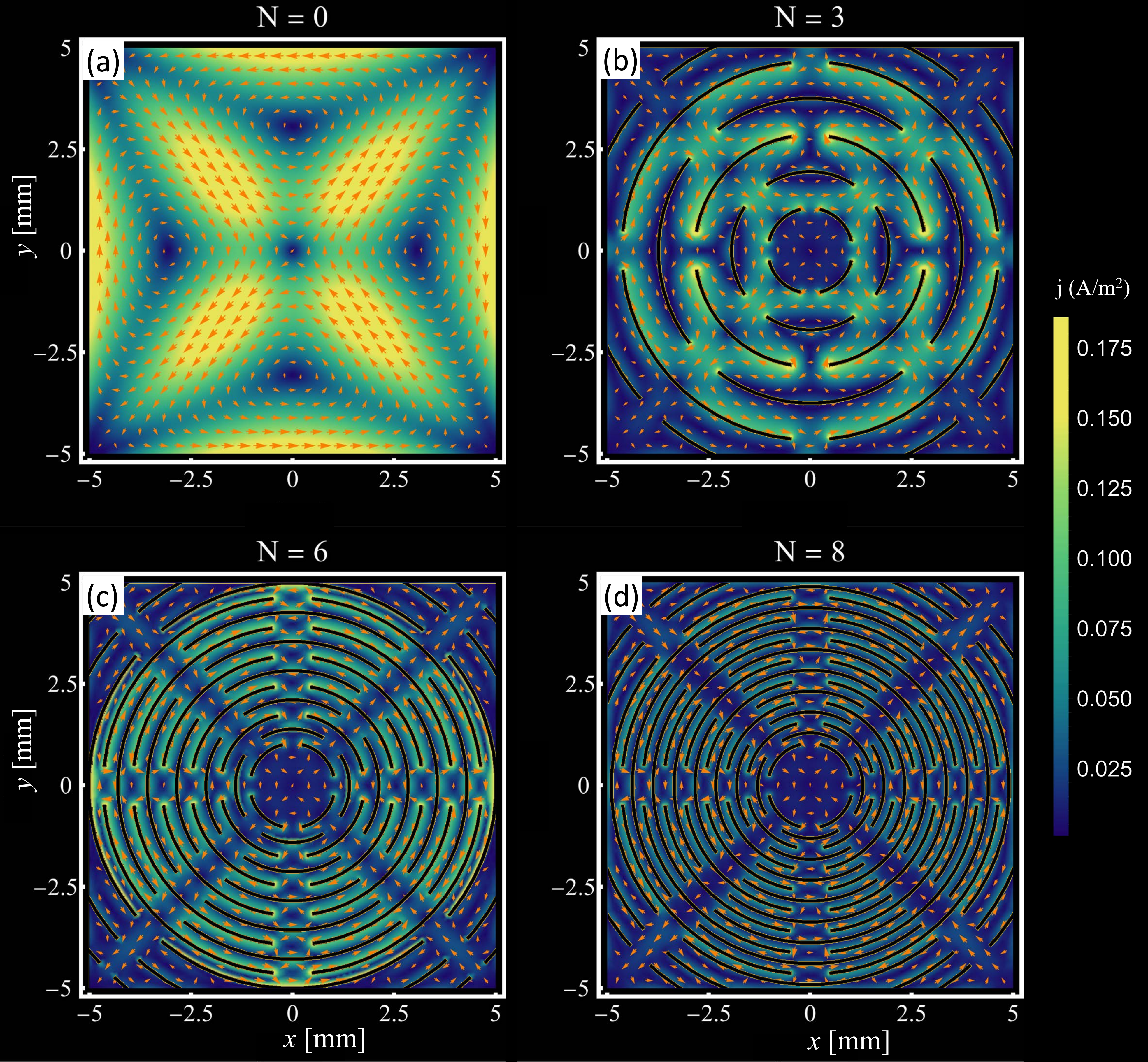}
\caption{Eddy currents simulated by COMSOL on the bottom surface of each plate. The plate is assumed to move vertically downwards away from its equilibrium position. The currents simulated by the two-dimensional Mathematica model are similar. We choose $v=-6\times10^{-6}\: {\rm m/s}$ for $n=0$ and $n=3$, and  $v=-12\times10^{-6}\:{\rm m/s}$ for $n=6$ and $n=8$ (the different velocities are to ensure the currents are still visible at higher $N$ values). Colour denotes the current magnitude, while arrows show the current direction. Around the edges of the slits the current can attain very large values not shown on our colour scale, but these occur in vanishingly small regions. As the number of slits increases, the eddy currents are significantly suppressed.} \label{fig:composite}
\end{figure*}

We also built a three-dimensional model using the commercial FEM package COMSOL. The eddy currents are plotted in \cref{fig:composite} (these do not significantly differ from those generated in Mathematica). Using this we were able to predict the absolute values of the quality factors , rather than simply their ratios. The results of the COMSOL simulation are also plotted in \cref{fig:qratios}, agreeing with the results from the experiment and Mathematica. 

The motional quality factor of the graphite plates was measured to increase as more slots were cut into the surface. This increase was consistent with both the two-dimensional Mathematica and three-dimensional COMSOL simulations of the eddy current damping. This indicates that the increase in quality factor is indeed due to suppression of eddy currents. Overall currents were suppressed by an order of magnitude, corresponding to an increase in quality factor of forty.

In \cref{fig:qratios} we can see that for $N=8$, the simulated quality factors are slightly lower than the experimental value. We attribute this primarily to a discrepancy between the designed slot patterns, and what is created by the femtosecond laser cutting. In \cref{fig:SlitPattern}, the slot patterns of the laser-cut samples are clearly wider than the CAD designs. Moreover, the machining process carves V-shaped slits which remove more graphite than expected, an effect which is is more pronounced at high slit density. Wider slots yield less eddy currents and thus reduced eddy damping, leading to higher quality factors in the actual samples than what is predicted by our models. 


Graphite is one of the strongest diamagnetic materials known, and has great potential for use in levitated technologies. However, due to its high electrical conductivity it exhibits strong eddy damping.
The ability to engineer this damping while retaining a strong diamagnetic susceptibility will permit researchers in a wide range of disciplines the ability to apply such conducting diamagnetic materials to situations where fast motional control is required. 
We show that by patterning the graphite plate with through-slots we can interrupt the eddy currents in a controlled manner, and gain detailed control over the eddy damping while retaining the strong diamagnetic lift. In this study we have not optimized the slotted pattern, and it is an interesting question whether one can produce designs which remove the least material, maintain the structural integrity of the plate, and control the eddy damping to the maximal extent. 

\section*{Acknowledgements}
This work was supported by the Okinawa Institute of Science and Technology Graduate University, Japan and Macquarie University, Sydney, Australia. The authors acknowledge technical assistance from E. Elerabi and P. Kennedy from the OIST Engineering Section. 
The graphite slab machining work was performed in part at the OptoFab node of the Australian National Fabrication Facility (ANFF) utilising Commonwealth and NSW State Government funding.

\section{Data Availability Statement}
The data and simulation codes that support the findings of this study are available from the corresponding author upon reasonable request.


\begin{thebibliography}{57}%
\makeatletter
\providecommand \@ifxundefined [1]{%
 \@ifx{#1\undefined}
}%
\providecommand \@ifnum [1]{%
 \ifnum #1\expandafter \@firstoftwo
 \else \expandafter \@secondoftwo
 \fi
}%
\providecommand \@ifx [1]{%
 \ifx #1\expandafter \@firstoftwo
 \else \expandafter \@secondoftwo
 \fi
}%
\providecommand \natexlab [1]{#1}%
\providecommand \enquote  [1]{``#1''}%
\providecommand \bibnamefont  [1]{#1}%
\providecommand \bibfnamefont [1]{#1}%
\providecommand \citenamefont [1]{#1}%
\providecommand \href@noop [0]{\@secondoftwo}%
\providecommand \href [0]{\begingroup \@sanitize@url \@href}%
\providecommand \@href[1]{\@@startlink{#1}\@@href}%
\providecommand \@@href[1]{\endgroup#1\@@endlink}%
\providecommand \@sanitize@url [0]{\catcode `\\12\catcode `\$12\catcode
  `\&12\catcode `\#12\catcode `\^12\catcode `\_12\catcode `\%12\relax}%
\providecommand \@@startlink[1]{}%
\providecommand \@@endlink[0]{}%
\providecommand \url  [0]{\begingroup\@sanitize@url \@url }%
\providecommand \@url [1]{\endgroup\@href {#1}{\urlprefix }}%
\providecommand \urlprefix  [0]{URL }%
\providecommand \Eprint [0]{\href }%
\providecommand \doibase [0]{https://doi.org/}%
\providecommand \selectlanguage [0]{\@gobble}%
\providecommand \bibinfo  [0]{\@secondoftwo}%
\providecommand \bibfield  [0]{\@secondoftwo}%
\providecommand \translation [1]{[#1]}%
\providecommand \BibitemOpen [0]{}%
\providecommand \bibitemStop [0]{}%
\providecommand \bibitemNoStop [0]{.\EOS\space}%
\providecommand \EOS [0]{\spacefactor3000\relax}%
\providecommand \BibitemShut  [1]{\csname bibitem#1\endcsname}%
\let\auto@bib@innerbib\@empty
\bibitem [{\citenamefont {Chen}\ \emph {et~al.}(2020)\citenamefont {Chen},
  \citenamefont {Ke{\c{s}}kekler}, \citenamefont {Alijani},\ and\ \citenamefont
  {Steeneken}}]{Chen2020}%
  \BibitemOpen
  \bibfield  {author} {\bibinfo {author} {\bibfnamefont {X.}~\bibnamefont
  {Chen}}, \bibinfo {author} {\bibfnamefont {A.}~\bibnamefont
  {Ke{\c{s}}kekler}}, \bibinfo {author} {\bibfnamefont {F.}~\bibnamefont
  {Alijani}},\ and\ \bibinfo {author} {\bibfnamefont {P.~G.}\ \bibnamefont
  {Steeneken}},\ }\bibfield  {title} {\bibinfo {title} {{Rigid body dynamics of
  diamagnetically levitating graphite resonators}},\ }\href
  {https://doi.org/10.1063/5.0009604} {\bibfield  {journal} {\bibinfo
  {journal} {Applied Physics Letters}\ }\textbf {\bibinfo {volume} {116}},\
  \bibinfo {pages} {243505} (\bibinfo {year} {2020})}\BibitemShut {NoStop}%
\bibitem [{\citenamefont {Berry}\ and\ \citenamefont {Geim}(1997)}]{Berry1997}%
  \BibitemOpen
  \bibfield  {author} {\bibinfo {author} {\bibfnamefont {M.~V.}\ \bibnamefont
  {Berry}}\ and\ \bibinfo {author} {\bibfnamefont {A.~K.}\ \bibnamefont
  {Geim}},\ }\bibfield  {title} {\bibinfo {title} {{Of flying frogs and
  levitrons}},\ }\href {https://doi.org/10.1088/0143-0807/18/4/012} {\bibfield
  {journal} {\bibinfo  {journal} {European Journal of Physics}\ }\textbf
  {\bibinfo {volume} {18}},\ \bibinfo {pages} {307} (\bibinfo {year}
  {1997})}\BibitemShut {NoStop}%
\bibitem [{\citenamefont {Simon}\ and\ \citenamefont {Geim}(2000)}]{Simon2000}%
  \BibitemOpen
  \bibfield  {author} {\bibinfo {author} {\bibfnamefont {M.~D.}\ \bibnamefont
  {Simon}}\ and\ \bibinfo {author} {\bibfnamefont {A.~K.}\ \bibnamefont
  {Geim}},\ }\bibfield  {title} {\bibinfo {title} {{Diamagnetic levitation:
  Flying frogs and floating magnets (invited)}},\ }\href
  {https://doi.org/10.1063/1.372654} {\bibfield  {journal} {\bibinfo  {journal}
  {Journal of Applied Physics}\ }\textbf {\bibinfo {volume} {87}},\ \bibinfo
  {pages} {6200} (\bibinfo {year} {2000})}\BibitemShut {NoStop}%
\bibitem [{\citenamefont {Simon}\ \emph {et~al.}(2001)\citenamefont {Simon},
  \citenamefont {Heflinger},\ and\ \citenamefont {Geim}}]{Simon2001}%
  \BibitemOpen
  \bibfield  {author} {\bibinfo {author} {\bibfnamefont {M.~D.}\ \bibnamefont
  {Simon}}, \bibinfo {author} {\bibfnamefont {L.~O.}\ \bibnamefont
  {Heflinger}},\ and\ \bibinfo {author} {\bibfnamefont {A.~K.}\ \bibnamefont
  {Geim}},\ }\bibfield  {title} {\bibinfo {title} {{Diamagnetically stabilized
  magnet levitation}},\ }\href {https://doi.org/10.1119/1.1375157} {\bibfield
  {journal} {\bibinfo  {journal} {American Journal of Physics}\ }\textbf
  {\bibinfo {volume} {69}},\ \bibinfo {pages} {702} (\bibinfo {year}
  {2001})}\BibitemShut {NoStop}%
\bibitem [{\citenamefont {Gonzalez-Ballestero}\ \emph
  {et~al.}(2021)\citenamefont {Gonzalez-Ballestero}, \citenamefont
  {Aspelmeyer}, \citenamefont {Novotny}, \citenamefont {Quidant},\ and\
  \citenamefont {Romero-Isart}}]{Gonzalez-Ballestero2021}%
  \BibitemOpen
  \bibfield  {author} {\bibinfo {author} {\bibfnamefont {C.}~\bibnamefont
  {Gonzalez-Ballestero}}, \bibinfo {author} {\bibfnamefont {M.}~\bibnamefont
  {Aspelmeyer}}, \bibinfo {author} {\bibfnamefont {L.}~\bibnamefont {Novotny}},
  \bibinfo {author} {\bibfnamefont {R.}~\bibnamefont {Quidant}},\ and\ \bibinfo
  {author} {\bibfnamefont {O.}~\bibnamefont {Romero-Isart}},\ }\bibfield
  {title} {\bibinfo {title} {{Levitodynamics: Levitation and control of
  microscopic objects in vacuum}},\ }\href
  {https://doi.org/10.1126/science.abg3027} {\bibfield  {journal} {\bibinfo
  {journal} {Science}\ }\textbf {\bibinfo {volume} {374}},\ \bibinfo {pages}
  {eabg3027} (\bibinfo {year} {2021})}\BibitemShut {NoStop}%
\bibitem [{\citenamefont {Gao}\ \emph {et~al.}(2021)\citenamefont {Gao},
  \citenamefont {Yan}, \citenamefont {Zou}, \citenamefont {Li}, \citenamefont
  {Peng}, \citenamefont {Meng},\ and\ \citenamefont
  {Zhang}}]{Gao2021MagneticProspects}%
  \BibitemOpen
  \bibfield  {author} {\bibinfo {author} {\bibfnamefont {Q.}~\bibnamefont
  {Gao}}, \bibinfo {author} {\bibfnamefont {H.}~\bibnamefont {Yan}}, \bibinfo
  {author} {\bibfnamefont {H.}~\bibnamefont {Zou}}, \bibinfo {author}
  {\bibfnamefont {W.}~\bibnamefont {Li}}, \bibinfo {author} {\bibfnamefont
  {Z.}~\bibnamefont {Peng}}, \bibinfo {author} {\bibfnamefont {G.}~\bibnamefont
  {Meng}},\ and\ \bibinfo {author} {\bibfnamefont {W.}~\bibnamefont {Zhang}},\
  }\bibfield  {title} {\bibinfo {title} {{Magnetic levitation using
  diamagnetism: Mechanism, applications and prospects}},\ }\href
  {https://doi.org/10.1007/s11431-020-1550-1} {\bibfield  {journal} {\bibinfo
  {journal} {Science China Technological Sciences}\ }\textbf {\bibinfo {volume}
  {64}},\ \bibinfo {pages} {44} (\bibinfo {year} {2021})}\BibitemShut {NoStop}%
\bibitem [{\citenamefont {Gieseler}\ and\ \citenamefont
  {Millen}(2018)}]{Gieseler2018}%
  \BibitemOpen
  \bibfield  {author} {\bibinfo {author} {\bibfnamefont {J.}~\bibnamefont
  {Gieseler}}\ and\ \bibinfo {author} {\bibfnamefont {J.}~\bibnamefont
  {Millen}},\ }\bibfield  {title} {\bibinfo {title} {{Levitated Nanoparticles
  for Microscopic Thermodynamicsâ€”A Review}},\ }\href
  {https://doi.org/10.3390/e20050326} {\bibfield  {journal} {\bibinfo
  {journal} {Entropy}\ }\textbf {\bibinfo {volume} {20}},\ \bibinfo {pages}
  {326} (\bibinfo {year} {2018})}\BibitemShut {NoStop}%
\bibitem [{\citenamefont {Pontin}\ \emph {et~al.}(2020)\citenamefont {Pontin},
  \citenamefont {Bullier}, \citenamefont {Toro{\v{s}}},\ and\ \citenamefont
  {Barker}}]{Pontin2020}%
  \BibitemOpen
  \bibfield  {author} {\bibinfo {author} {\bibfnamefont {A.}~\bibnamefont
  {Pontin}}, \bibinfo {author} {\bibfnamefont {N.~P.}\ \bibnamefont {Bullier}},
  \bibinfo {author} {\bibfnamefont {M.}~\bibnamefont {Toro{\v{s}}}},\ and\
  \bibinfo {author} {\bibfnamefont {P.~F.}\ \bibnamefont {Barker}},\ }\bibfield
   {title} {\bibinfo {title} {{Ultranarrow-linewidth levitated nano-oscillator
  for testing dissipative wave-function collapse}},\ }\href
  {https://doi.org/10.1103/PhysRevResearch.2.023349} {\bibfield  {journal}
  {\bibinfo  {journal} {Physical Review Research}\ }\textbf {\bibinfo {volume}
  {2}},\ \bibinfo {pages} {023349} (\bibinfo {year} {2020})}\BibitemShut
  {NoStop}%
\bibitem [{\citenamefont {Moore}\ and\ \citenamefont
  {Geraci}(2021)}]{Moore2021}%
  \BibitemOpen
  \bibfield  {author} {\bibinfo {author} {\bibfnamefont {D.~C.}\ \bibnamefont
  {Moore}}\ and\ \bibinfo {author} {\bibfnamefont {A.~A.}\ \bibnamefont
  {Geraci}},\ }\bibfield  {title} {\bibinfo {title} {{Searching for new physics
  using optically levitated sensors}},\ }\href
  {https://doi.org/10.1088/2058-9565/abcf8a} {\bibfield  {journal} {\bibinfo
  {journal} {Quantum Science and Technology}\ }\textbf {\bibinfo {volume}
  {6}},\ \bibinfo {pages} {014008} (\bibinfo {year} {2021})}\BibitemShut
  {NoStop}%
\bibitem [{\citenamefont {Xiong}\ \emph {et~al.}(2021)\citenamefont {Xiong},
  \citenamefont {Wu}, \citenamefont {Leng}, \citenamefont {Li}, \citenamefont
  {Duan}, \citenamefont {Kong}, \citenamefont {Huang}, \citenamefont {Li},
  \citenamefont {Gao}, \citenamefont {Rong},\ and\ \citenamefont
  {Du}}]{Xiong2021SearchingSensor}%
  \BibitemOpen
  \bibfield  {author} {\bibinfo {author} {\bibfnamefont {F.}~\bibnamefont
  {Xiong}}, \bibinfo {author} {\bibfnamefont {T.}~\bibnamefont {Wu}}, \bibinfo
  {author} {\bibfnamefont {Y.}~\bibnamefont {Leng}}, \bibinfo {author}
  {\bibfnamefont {R.}~\bibnamefont {Li}}, \bibinfo {author} {\bibfnamefont
  {C.-K.}\ \bibnamefont {Duan}}, \bibinfo {author} {\bibfnamefont
  {X.}~\bibnamefont {Kong}}, \bibinfo {author} {\bibfnamefont {P.}~\bibnamefont
  {Huang}}, \bibinfo {author} {\bibfnamefont {Z.}~\bibnamefont {Li}}, \bibinfo
  {author} {\bibfnamefont {Y.}~\bibnamefont {Gao}}, \bibinfo {author}
  {\bibfnamefont {X.}~\bibnamefont {Rong}},\ and\ \bibinfo {author}
  {\bibfnamefont {J.}~\bibnamefont {Du}},\ }\bibfield  {title} {\bibinfo
  {title} {{Searching spin-mass interaction using a diamagnetic levitated
  magnetic-resonance force sensor}},\ }\href
  {https://doi.org/10.1103/PhysRevResearch.3.013205} {\bibfield  {journal}
  {\bibinfo  {journal} {Physical Review Research}\ }\textbf {\bibinfo {volume}
  {3}},\ \bibinfo {pages} {013205} (\bibinfo {year} {2021})}\BibitemShut
  {NoStop}%
\bibitem [{\citenamefont {Timberlake}\ \emph {et~al.}(2021)\citenamefont
  {Timberlake}, \citenamefont {Vinante}, \citenamefont {Shankar}, \citenamefont
  {Lapi},\ and\ \citenamefont {Ulbricht}}]{Timberlake2021ProbingResonators}%
  \BibitemOpen
  \bibfield  {author} {\bibinfo {author} {\bibfnamefont {C.}~\bibnamefont
  {Timberlake}}, \bibinfo {author} {\bibfnamefont {A.}~\bibnamefont {Vinante}},
  \bibinfo {author} {\bibfnamefont {F.}~\bibnamefont {Shankar}}, \bibinfo
  {author} {\bibfnamefont {A.}~\bibnamefont {Lapi}},\ and\ \bibinfo {author}
  {\bibfnamefont {H.}~\bibnamefont {Ulbricht}},\ }\bibfield  {title} {\bibinfo
  {title} {{Probing modified gravity with magnetically levitated resonators}},\
  }\href {https://doi.org/10.1103/PhysRevD.104.L101101} {\bibfield  {journal}
  {\bibinfo  {journal} {Physical Review D}\ }\textbf {\bibinfo {volume}
  {104}},\ \bibinfo {pages} {L101101} (\bibinfo {year} {2021})}\BibitemShut
  {NoStop}%
\bibitem [{\citenamefont {Bose}\ \emph {et~al.}(2017)\citenamefont {Bose},
  \citenamefont {Mazumdar}, \citenamefont {Morley}, \citenamefont {Ulbricht},
  \citenamefont {Toro{\v{s}}}, \citenamefont {Paternostro}, \citenamefont
  {Geraci}, \citenamefont {Barker}, \citenamefont {Kim},\ and\ \citenamefont
  {Milburn}}]{Bose2017}%
  \BibitemOpen
  \bibfield  {author} {\bibinfo {author} {\bibfnamefont {S.}~\bibnamefont
  {Bose}}, \bibinfo {author} {\bibfnamefont {A.}~\bibnamefont {Mazumdar}},
  \bibinfo {author} {\bibfnamefont {G.~W.}\ \bibnamefont {Morley}}, \bibinfo
  {author} {\bibfnamefont {H.}~\bibnamefont {Ulbricht}}, \bibinfo {author}
  {\bibfnamefont {M.}~\bibnamefont {Toro{\v{s}}}}, \bibinfo {author}
  {\bibfnamefont {M.}~\bibnamefont {Paternostro}}, \bibinfo {author}
  {\bibfnamefont {A.~A.}\ \bibnamefont {Geraci}}, \bibinfo {author}
  {\bibfnamefont {P.~F.}\ \bibnamefont {Barker}}, \bibinfo {author}
  {\bibfnamefont {M.~S.}\ \bibnamefont {Kim}},\ and\ \bibinfo {author}
  {\bibfnamefont {G.}~\bibnamefont {Milburn}},\ }\bibfield  {title} {\bibinfo
  {title} {{Spin Entanglement Witness for Quantum Gravity}},\ }\href
  {https://doi.org/10.1103/PhysRevLett.119.240401} {\bibfield  {journal}
  {\bibinfo  {journal} {Physical Review Letters}\ }\textbf {\bibinfo {volume}
  {119}},\ \bibinfo {pages} {240401} (\bibinfo {year} {2017})}\BibitemShut
  {NoStop}%
\bibitem [{\citenamefont {Marletto}\ and\ \citenamefont
  {Vedral}(2017)}]{Marletto2017}%
  \BibitemOpen
  \bibfield  {author} {\bibinfo {author} {\bibfnamefont {C.}~\bibnamefont
  {Marletto}}\ and\ \bibinfo {author} {\bibfnamefont {V.}~\bibnamefont
  {Vedral}},\ }\bibfield  {title} {\bibinfo {title} {{Gravitationally Induced
  Entanglement between Two Massive Particles is Sufficient Evidence of Quantum
  Effects in Gravity}},\ }\href
  {https://doi.org/10.1103/PhysRevLett.119.240402} {\bibfield  {journal}
  {\bibinfo  {journal} {Physical Review Letters}\ }\textbf {\bibinfo {volume}
  {119}},\ \bibinfo {pages} {240402} (\bibinfo {year} {2017})}\BibitemShut
  {NoStop}%
\bibitem [{\citenamefont {Christodoulou}\ and\ \citenamefont
  {Rovelli}(2019)}]{Christodoulou2019}%
  \BibitemOpen
  \bibfield  {author} {\bibinfo {author} {\bibfnamefont {M.}~\bibnamefont
  {Christodoulou}}\ and\ \bibinfo {author} {\bibfnamefont {C.}~\bibnamefont
  {Rovelli}},\ }\bibfield  {title} {\bibinfo {title} {{On the possibility of
  laboratory evidence for quantum superposition of geometries}},\ }\href
  {https://doi.org/10.1016/j.physletb.2019.03.015} {\bibfield  {journal}
  {\bibinfo  {journal} {Physics Letters, Section B: Nuclear, Elementary
  Particle and High-Energy Physics}\ }\textbf {\bibinfo {volume} {792}},\
  \bibinfo {pages} {64} (\bibinfo {year} {2019})}\BibitemShut {NoStop}%
\bibitem [{\citenamefont {Carlesso}\ \emph {et~al.}(2019)\citenamefont
  {Carlesso}, \citenamefont {Bassi}, \citenamefont {Paternostro},\ and\
  \citenamefont {Ulbricht}}]{Carlesso2019}%
  \BibitemOpen
  \bibfield  {author} {\bibinfo {author} {\bibfnamefont {M.}~\bibnamefont
  {Carlesso}}, \bibinfo {author} {\bibfnamefont {A.}~\bibnamefont {Bassi}},
  \bibinfo {author} {\bibfnamefont {M.}~\bibnamefont {Paternostro}},\ and\
  \bibinfo {author} {\bibfnamefont {H.}~\bibnamefont {Ulbricht}},\ }\bibfield
  {title} {\bibinfo {title} {{Testing the gravitational field generated by a
  quantum superposition}},\ }\href {https://doi.org/10.1088/1367-2630/ab41c1}
  {\bibfield  {journal} {\bibinfo  {journal} {New Journal of Physics}\ }\textbf
  {\bibinfo {volume} {21}},\ \bibinfo {pages} {093052} (\bibinfo {year}
  {2019})}\BibitemShut {NoStop}%
\bibitem [{\citenamefont {Nguyen}\ and\ \citenamefont
  {Bernards}(2020)}]{Nguyen2020}%
  \BibitemOpen
  \bibfield  {author} {\bibinfo {author} {\bibfnamefont {H.~C.}\ \bibnamefont
  {Nguyen}}\ and\ \bibinfo {author} {\bibfnamefont {F.}~\bibnamefont
  {Bernards}},\ }\bibfield  {title} {\bibinfo {title} {{Entanglement dynamics
  of two mesoscopic objects with gravitational interaction}},\ }\href
  {https://doi.org/10.1140/epjd/e2020-10077-8} {\bibfield  {journal} {\bibinfo
  {journal} {European Physical Journal D}\ }\textbf {\bibinfo {volume} {74}},\
  \bibinfo {pages} {2} (\bibinfo {year} {2020})}\BibitemShut {NoStop}%
\bibitem [{\citenamefont {Marshman}\ \emph {et~al.}(2022)\citenamefont
  {Marshman}, \citenamefont {Mazumdar}, \citenamefont {Folman},\ and\
  \citenamefont {Bose}}]{Marshman2022ConstructingInterferometer}%
  \BibitemOpen
  \bibfield  {author} {\bibinfo {author} {\bibfnamefont {R.~J.}\ \bibnamefont
  {Marshman}}, \bibinfo {author} {\bibfnamefont {A.}~\bibnamefont {Mazumdar}},
  \bibinfo {author} {\bibfnamefont {R.}~\bibnamefont {Folman}},\ and\ \bibinfo
  {author} {\bibfnamefont {S.}~\bibnamefont {Bose}},\ }\bibfield  {title}
  {\bibinfo {title} {{Constructing nano-object quantum superpositions with a
  Stern-Gerlach interferometer}},\ }\href
  {https://doi.org/10.1103/PhysRevResearch.4.023087} {\bibfield  {journal}
  {\bibinfo  {journal} {Physical Review Research}\ }\textbf {\bibinfo {volume}
  {4}},\ \bibinfo {pages} {023087} (\bibinfo {year} {2022})}\BibitemShut
  {NoStop}%
\bibitem [{\citenamefont {Garmire}\ \emph {et~al.}(2007)\citenamefont
  {Garmire}, \citenamefont {Choo}, \citenamefont {Kant}, \citenamefont
  {Govindjee}, \citenamefont {S{\'{e}}quin}, \citenamefont {Muller},\ and\
  \citenamefont {Demmel}}]{Garmire2007}%
  \BibitemOpen
  \bibfield  {author} {\bibinfo {author} {\bibfnamefont {D.}~\bibnamefont
  {Garmire}}, \bibinfo {author} {\bibfnamefont {H.}~\bibnamefont {Choo}},
  \bibinfo {author} {\bibfnamefont {R.}~\bibnamefont {Kant}}, \bibinfo {author}
  {\bibfnamefont {S.}~\bibnamefont {Govindjee}}, \bibinfo {author}
  {\bibfnamefont {C.~H.}\ \bibnamefont {S{\'{e}}quin}}, \bibinfo {author}
  {\bibfnamefont {R.~S.}\ \bibnamefont {Muller}},\ and\ \bibinfo {author}
  {\bibfnamefont {J.}~\bibnamefont {Demmel}},\ }\bibfield  {title} {\bibinfo
  {title} {{Diamagnetically levitated MEMS accelerometers}},\ }in\ \href
  {https://doi.org/10.1109/SENSOR.2007.4300352} {\emph {\bibinfo {booktitle}
  {TRANSDUCERS and EUROSENSORS '07 - 4th International Conference on
  Solid-State Sensors, Actuators and Microsystems}}}\ (\bibinfo {year} {2007})\
  pp.\ \bibinfo {pages} {1203--1206}\BibitemShut {NoStop}%
\bibitem [{\citenamefont {Hempston}\ \emph {et~al.}(2017)\citenamefont
  {Hempston}, \citenamefont {Vovrosh}, \citenamefont {Toro{\v{s}}},
  \citenamefont {Winstone}, \citenamefont {Rashid},\ and\ \citenamefont
  {Ulbricht}}]{Hempston2017}%
  \BibitemOpen
  \bibfield  {author} {\bibinfo {author} {\bibfnamefont {D.}~\bibnamefont
  {Hempston}}, \bibinfo {author} {\bibfnamefont {J.}~\bibnamefont {Vovrosh}},
  \bibinfo {author} {\bibfnamefont {M.}~\bibnamefont {Toro{\v{s}}}}, \bibinfo
  {author} {\bibfnamefont {G.}~\bibnamefont {Winstone}}, \bibinfo {author}
  {\bibfnamefont {M.}~\bibnamefont {Rashid}},\ and\ \bibinfo {author}
  {\bibfnamefont {H.}~\bibnamefont {Ulbricht}},\ }\bibfield  {title} {\bibinfo
  {title} {{Force sensing with an optically levitated charged nanoparticle}},\
  }\href {https://doi.org/10.1063/1.4993555} {\bibfield  {journal} {\bibinfo
  {journal} {Applied Physics Letters}\ }\textbf {\bibinfo {volume} {111}},\
  \bibinfo {pages} {133111} (\bibinfo {year} {2017})}\BibitemShut {NoStop}%
\bibitem [{\citenamefont {Prat-Camps}\ \emph {et~al.}(2017)\citenamefont
  {Prat-Camps}, \citenamefont {Teo}, \citenamefont {Rusconi}, \citenamefont
  {Wieczorek},\ and\ \citenamefont {Romero-Isart}}]{Prat-camps2017}%
  \BibitemOpen
  \bibfield  {author} {\bibinfo {author} {\bibfnamefont {J.}~\bibnamefont
  {Prat-Camps}}, \bibinfo {author} {\bibfnamefont {C.}~\bibnamefont {Teo}},
  \bibinfo {author} {\bibfnamefont {C.~C.}\ \bibnamefont {Rusconi}}, \bibinfo
  {author} {\bibfnamefont {W.}~\bibnamefont {Wieczorek}},\ and\ \bibinfo
  {author} {\bibfnamefont {O.}~\bibnamefont {Romero-Isart}},\ }\bibfield
  {title} {\bibinfo {title} {{Ultrasensitive Inertial and Force Sensors with
  Diamagnetically Levitated Magnets}},\ }\href
  {https://doi.org/10.1103/PhysRevApplied.8.034002} {\bibfield  {journal}
  {\bibinfo  {journal} {Physical Review Applied}\ }\textbf {\bibinfo {volume}
  {8}},\ \bibinfo {pages} {034002} (\bibinfo {year} {2017})}\BibitemShut
  {NoStop}%
\bibitem [{\citenamefont {Timberlake}\ \emph {et~al.}(2019)\citenamefont
  {Timberlake}, \citenamefont {Gasbarri}, \citenamefont {Vinante},
  \citenamefont {Setter},\ and\ \citenamefont {Ulbricht}}]{Timberlake2019}%
  \BibitemOpen
  \bibfield  {author} {\bibinfo {author} {\bibfnamefont {C.}~\bibnamefont
  {Timberlake}}, \bibinfo {author} {\bibfnamefont {G.}~\bibnamefont
  {Gasbarri}}, \bibinfo {author} {\bibfnamefont {A.}~\bibnamefont {Vinante}},
  \bibinfo {author} {\bibfnamefont {A.}~\bibnamefont {Setter}},\ and\ \bibinfo
  {author} {\bibfnamefont {H.}~\bibnamefont {Ulbricht}},\ }\bibfield  {title}
  {\bibinfo {title} {{Acceleration sensing with magnetically levitated
  oscillators above a superconductor}},\ }\href
  {https://doi.org/10.1063/1.5129145} {\bibfield  {journal} {\bibinfo
  {journal} {Applied Physics Letters}\ }\textbf {\bibinfo {volume} {115}},\
  \bibinfo {pages} {224101} (\bibinfo {year} {2019})}\BibitemShut {NoStop}%
\bibitem [{\citenamefont {Monteiro}\ \emph {et~al.}(2020)\citenamefont
  {Monteiro}, \citenamefont {Li}, \citenamefont {Afek}, \citenamefont {Li},
  \citenamefont {Mossman},\ and\ \citenamefont {Moore}}]{Monteiro2020}%
  \BibitemOpen
  \bibfield  {author} {\bibinfo {author} {\bibfnamefont {F.}~\bibnamefont
  {Monteiro}}, \bibinfo {author} {\bibfnamefont {W.}~\bibnamefont {Li}},
  \bibinfo {author} {\bibfnamefont {G.}~\bibnamefont {Afek}}, \bibinfo {author}
  {\bibfnamefont {C.~L.}\ \bibnamefont {Li}}, \bibinfo {author} {\bibfnamefont
  {M.}~\bibnamefont {Mossman}},\ and\ \bibinfo {author} {\bibfnamefont {D.~C.}\
  \bibnamefont {Moore}},\ }\bibfield  {title} {\bibinfo {title} {{Force and
  acceleration sensing with optically levitated nanogram masses at microkelvin
  temperatures}},\ }\href {https://doi.org/10.1103/PhysRevA.101.053835}
  {\bibfield  {journal} {\bibinfo  {journal} {Physical Review A}\ }\textbf
  {\bibinfo {volume} {101}},\ \bibinfo {pages} {53835} (\bibinfo {year}
  {2020})}\BibitemShut {NoStop}%
\bibitem [{\citenamefont {Lewandowski}\ \emph
  {et~al.}(2021{\natexlab{a}})\citenamefont {Lewandowski}, \citenamefont
  {Knowles}, \citenamefont {Etienne},\ and\ \citenamefont
  {Dâ€™Urso}}]{Lewandowski2021}%
  \BibitemOpen
  \bibfield  {author} {\bibinfo {author} {\bibfnamefont {C.~W.}\ \bibnamefont
  {Lewandowski}}, \bibinfo {author} {\bibfnamefont {T.~D.}\ \bibnamefont
  {Knowles}}, \bibinfo {author} {\bibfnamefont {Z.~B.}\ \bibnamefont
  {Etienne}},\ and\ \bibinfo {author} {\bibfnamefont {B.}~\bibnamefont
  {Dâ€™Urso}},\ }\bibfield  {title} {\bibinfo {title} {{High-Sensitivity
  Accelerometry with a Feedback-Cooled Magnetically Levitated Microsphere}},\
  }\href {https://doi.org/10.1103/PhysRevApplied.15.014050} {\bibfield
  {journal} {\bibinfo  {journal} {Physical Review Applied}\ }\textbf {\bibinfo
  {volume} {15}},\ \bibinfo {pages} {014050} (\bibinfo {year}
  {2021}{\natexlab{a}})}\BibitemShut {NoStop}%
\bibitem [{\citenamefont {Arvanitaki}\ and\ \citenamefont
  {Geraci}(2013)}]{Arvanitaki2013DetectingSensors}%
  \BibitemOpen
  \bibfield  {author} {\bibinfo {author} {\bibfnamefont {A.}~\bibnamefont
  {Arvanitaki}}\ and\ \bibinfo {author} {\bibfnamefont {A.~A.}\ \bibnamefont
  {Geraci}},\ }\bibfield  {title} {\bibinfo {title} {{Detecting High-Frequency
  Gravitational Waves with Optically Levitated Sensors}},\ }\href
  {https://doi.org/10.1103/PhysRevLett.110.071105} {\bibfield  {journal}
  {\bibinfo  {journal} {Physical Review Letters}\ }\textbf {\bibinfo {volume}
  {110}},\ \bibinfo {pages} {071105} (\bibinfo {year} {2013})}\BibitemShut
  {NoStop}%
\bibitem [{\citenamefont {Yin}\ \emph {et~al.}(2022)\citenamefont {Yin},
  \citenamefont {Li}, \citenamefont {Yin}, \citenamefont {Xu}, \citenamefont
  {Bian}, \citenamefont {Xie}, \citenamefont {Duan}, \citenamefont {Huang},
  \citenamefont {He},\ and\ \citenamefont {Du}}]{Yin2022ExperimentsEnergyc}%
  \BibitemOpen
  \bibfield  {author} {\bibinfo {author} {\bibfnamefont {P.}~\bibnamefont
  {Yin}}, \bibinfo {author} {\bibfnamefont {R.}~\bibnamefont {Li}}, \bibinfo
  {author} {\bibfnamefont {C.}~\bibnamefont {Yin}}, \bibinfo {author}
  {\bibfnamefont {X.}~\bibnamefont {Xu}}, \bibinfo {author} {\bibfnamefont
  {X.}~\bibnamefont {Bian}}, \bibinfo {author} {\bibfnamefont {H.}~\bibnamefont
  {Xie}}, \bibinfo {author} {\bibfnamefont {C.-K.}\ \bibnamefont {Duan}},
  \bibinfo {author} {\bibfnamefont {P.}~\bibnamefont {Huang}}, \bibinfo
  {author} {\bibfnamefont {J.-h.}\ \bibnamefont {He}},\ and\ \bibinfo {author}
  {\bibfnamefont {J.}~\bibnamefont {Du}},\ }\bibfield  {title} {\bibinfo
  {title} {{Experiments with levitated force sensor challenge theories of dark
  energy}},\ }\bibfield  {journal} {\bibinfo  {journal} {Nature Physics}\
  }\href {https://doi.org/10.1038/s41567-022-01706-9}
  {10.1038/s41567-022-01706-9} (\bibinfo {year} {2022})\BibitemShut {NoStop}%
\bibitem [{\citenamefont {Kumar}\ and\ \citenamefont
  {Bhattacharya}(2017)}]{Kumar2017}%
  \BibitemOpen
  \bibfield  {author} {\bibinfo {author} {\bibfnamefont {P.}~\bibnamefont
  {Kumar}}\ and\ \bibinfo {author} {\bibfnamefont {M.}~\bibnamefont
  {Bhattacharya}},\ }\bibfield  {title} {\bibinfo {title} {{Magnetometry via
  spin-mechanical coupling in levitated optomechanics}},\ }\href
  {https://doi.org/10.1364/OE.25.019568} {\bibfield  {journal} {\bibinfo
  {journal} {Optics Express}\ }\textbf {\bibinfo {volume} {25}},\ \bibinfo
  {pages} {19568} (\bibinfo {year} {2017})}\BibitemShut {NoStop}%
\bibitem [{\citenamefont {Chen}\ \emph {et~al.}(2021)\citenamefont {Chen},
  \citenamefont {Kothari}, \citenamefont {Ke{\c{s}}kekler}, \citenamefont
  {Steeneken},\ and\ \citenamefont {Alijani}}]{Chen2021DiamagneticallyScaleb}%
  \BibitemOpen
  \bibfield  {author} {\bibinfo {author} {\bibfnamefont {X.}~\bibnamefont
  {Chen}}, \bibinfo {author} {\bibfnamefont {N.}~\bibnamefont {Kothari}},
  \bibinfo {author} {\bibfnamefont {A.}~\bibnamefont {Ke{\c{s}}kekler}},
  \bibinfo {author} {\bibfnamefont {P.~G.}\ \bibnamefont {Steeneken}},\ and\
  \bibinfo {author} {\bibfnamefont {F.}~\bibnamefont {Alijani}},\ }\bibfield
  {title} {\bibinfo {title} {{Diamagnetically levitating resonant weighing
  scale}},\ }\href {https://doi.org/10.1016/j.sna.2021.112842} {\bibfield
  {journal} {\bibinfo  {journal} {Sensors and Actuators A: Physical}\ }\textbf
  {\bibinfo {volume} {330}},\ \bibinfo {pages} {112842} (\bibinfo {year}
  {2021})}\BibitemShut {NoStop}%
\bibitem [{\citenamefont {Vaskuri}\ \emph {et~al.}(2021)\citenamefont
  {Vaskuri}, \citenamefont {Rahn}, \citenamefont {Williams},\ and\
  \citenamefont {Lehman}}]{Vaskuri2021AbsoluteMass}%
  \BibitemOpen
  \bibfield  {author} {\bibinfo {author} {\bibfnamefont {A.~K.}\ \bibnamefont
  {Vaskuri}}, \bibinfo {author} {\bibfnamefont {D.~W.}\ \bibnamefont {Rahn}},
  \bibinfo {author} {\bibfnamefont {P.~A.}\ \bibnamefont {Williams}},\ and\
  \bibinfo {author} {\bibfnamefont {J.~H.}\ \bibnamefont {Lehman}},\ }\bibfield
   {title} {\bibinfo {title} {{Absolute radiation pressure detector using a
  diamagnetically levitating test mass}},\ }\href
  {https://doi.org/10.1364/optica.437586} {\bibfield  {journal} {\bibinfo
  {journal} {Optica}\ }\textbf {\bibinfo {volume} {8}},\ \bibinfo {pages}
  {1380} (\bibinfo {year} {2021})}\BibitemShut {NoStop}%
\bibitem [{\citenamefont {Tebbenjohanns}\ \emph {et~al.}(2020)\citenamefont
  {Tebbenjohanns}, \citenamefont {Frimmer}, \citenamefont {Jain}, \citenamefont
  {Windey},\ and\ \citenamefont {Novotny}}]{Tebbenjohanns2020MotionalSpace}%
  \BibitemOpen
  \bibfield  {author} {\bibinfo {author} {\bibfnamefont {F.}~\bibnamefont
  {Tebbenjohanns}}, \bibinfo {author} {\bibfnamefont {M.}~\bibnamefont
  {Frimmer}}, \bibinfo {author} {\bibfnamefont {V.}~\bibnamefont {Jain}},
  \bibinfo {author} {\bibfnamefont {D.}~\bibnamefont {Windey}},\ and\ \bibinfo
  {author} {\bibfnamefont {L.}~\bibnamefont {Novotny}},\ }\bibfield  {title}
  {\bibinfo {title} {{Motional Sideband Asymmetry of a Nanoparticle Optically
  Levitated in Free Space}},\ }\href
  {https://doi.org/10.1103/PhysRevLett.124.013603} {\bibfield  {journal}
  {\bibinfo  {journal} {Physical Review Letters}\ }\textbf {\bibinfo {volume}
  {124}},\ \bibinfo {pages} {13603} (\bibinfo {year} {2020})}\BibitemShut
  {NoStop}%
\bibitem [{\citenamefont {Deli{\'{c}}}\ \emph {et~al.}(2020)\citenamefont
  {Deli{\'{c}}}, \citenamefont {Reisenbauer}, \citenamefont {Dare},
  \citenamefont {Grass}, \citenamefont {Vuleti{\'{c}}}, \citenamefont
  {Kiesel},\ and\ \citenamefont {Aspelmeyer}}]{Delic2020CoolingState}%
  \BibitemOpen
  \bibfield  {author} {\bibinfo {author} {\bibfnamefont {U.}~\bibnamefont
  {Deli{\'{c}}}}, \bibinfo {author} {\bibfnamefont {M.}~\bibnamefont
  {Reisenbauer}}, \bibinfo {author} {\bibfnamefont {K.}~\bibnamefont {Dare}},
  \bibinfo {author} {\bibfnamefont {D.}~\bibnamefont {Grass}}, \bibinfo
  {author} {\bibfnamefont {V.}~\bibnamefont {Vuleti{\'{c}}}}, \bibinfo {author}
  {\bibfnamefont {N.}~\bibnamefont {Kiesel}},\ and\ \bibinfo {author}
  {\bibfnamefont {M.}~\bibnamefont {Aspelmeyer}},\ }\bibfield  {title}
  {\bibinfo {title} {{Cooling of a levitated nanoparticle to the motional
  quantum ground state}},\ }\href {https://doi.org/10.1126/science.aba3993}
  {\bibfield  {journal} {\bibinfo  {journal} {Science}\ }\textbf {\bibinfo
  {volume} {367}},\ \bibinfo {pages} {892} (\bibinfo {year}
  {2020})}\BibitemShut {NoStop}%
\bibitem [{\citenamefont {Magrini}\ \emph {et~al.}(2021)\citenamefont
  {Magrini}, \citenamefont {Rosenzweig}, \citenamefont {Bach}, \citenamefont
  {Deutschmann-Olek}, \citenamefont {Hofer}, \citenamefont {Hong},
  \citenamefont {Kiesel}, \citenamefont {Kugi},\ and\ \citenamefont
  {Aspelmeyer}}]{Magrini2021Real-timeTemperature}%
  \BibitemOpen
  \bibfield  {author} {\bibinfo {author} {\bibfnamefont {L.}~\bibnamefont
  {Magrini}}, \bibinfo {author} {\bibfnamefont {P.}~\bibnamefont {Rosenzweig}},
  \bibinfo {author} {\bibfnamefont {C.}~\bibnamefont {Bach}}, \bibinfo {author}
  {\bibfnamefont {A.}~\bibnamefont {Deutschmann-Olek}}, \bibinfo {author}
  {\bibfnamefont {S.~G.}\ \bibnamefont {Hofer}}, \bibinfo {author}
  {\bibfnamefont {S.}~\bibnamefont {Hong}}, \bibinfo {author} {\bibfnamefont
  {N.}~\bibnamefont {Kiesel}}, \bibinfo {author} {\bibfnamefont
  {A.}~\bibnamefont {Kugi}},\ and\ \bibinfo {author} {\bibfnamefont
  {M.}~\bibnamefont {Aspelmeyer}},\ }\bibfield  {title} {\bibinfo {title}
  {{Real-time optimal quantum control of mechanical motion at room
  temperature}},\ }\href {https://doi.org/10.1038/s41586-021-03602-3}
  {\bibfield  {journal} {\bibinfo  {journal} {Nature}\ }\textbf {\bibinfo
  {volume} {595}},\ \bibinfo {pages} {373} (\bibinfo {year}
  {2021})}\BibitemShut {NoStop}%
\bibitem [{\citenamefont {Tebbenjohanns}\ \emph {et~al.}(2021)\citenamefont
  {Tebbenjohanns}, \citenamefont {Mattana}, \citenamefont {Rossi},
  \citenamefont {Frimmer},\ and\ \citenamefont
  {Novotny}}]{Tebbenjohanns2021QuantumSpace}%
  \BibitemOpen
  \bibfield  {author} {\bibinfo {author} {\bibfnamefont {F.}~\bibnamefont
  {Tebbenjohanns}}, \bibinfo {author} {\bibfnamefont {M.~L.}\ \bibnamefont
  {Mattana}}, \bibinfo {author} {\bibfnamefont {M.}~\bibnamefont {Rossi}},
  \bibinfo {author} {\bibfnamefont {M.}~\bibnamefont {Frimmer}},\ and\ \bibinfo
  {author} {\bibfnamefont {L.}~\bibnamefont {Novotny}},\ }\bibfield  {title}
  {\bibinfo {title} {{Quantum control of a nanoparticle optically levitated in
  cryogenic free space}},\ }\href {https://doi.org/10.1038/s41586-021-03617-w}
  {\bibfield  {journal} {\bibinfo  {journal} {Nature}\ }\textbf {\bibinfo
  {volume} {595}},\ \bibinfo {pages} {378} (\bibinfo {year}
  {2021})}\BibitemShut {NoStop}%
\bibitem [{\citenamefont {Millen}\ and\ \citenamefont
  {Stickler}(2020)}]{Millen2020QuantumParticles}%
  \BibitemOpen
  \bibfield  {author} {\bibinfo {author} {\bibfnamefont {J.}~\bibnamefont
  {Millen}}\ and\ \bibinfo {author} {\bibfnamefont {B.~A.}\ \bibnamefont
  {Stickler}},\ }\bibfield  {title} {\bibinfo {title} {{Quantum experiments
  with microscale particles}},\ }\href
  {https://doi.org/10.1080/00107514.2020.1854497} {\bibfield  {journal}
  {\bibinfo  {journal} {Contemporary Physics}\ }\textbf {\bibinfo {volume}
  {61}},\ \bibinfo {pages} {155} (\bibinfo {year} {2020})}\BibitemShut
  {NoStop}%
\bibitem [{\citenamefont {Pelrine}\ \emph {et~al.}(2012)\citenamefont
  {Pelrine}, \citenamefont {Wong-Foy}, \citenamefont {McCoy}, \citenamefont
  {Holeman}, \citenamefont {Mahoney}, \citenamefont {Myers}, \citenamefont
  {Herson},\ and\ \citenamefont {Low}}]{Pelrine2012DiamagneticallyProperties}%
  \BibitemOpen
  \bibfield  {author} {\bibinfo {author} {\bibfnamefont {R.}~\bibnamefont
  {Pelrine}}, \bibinfo {author} {\bibfnamefont {A.}~\bibnamefont {Wong-Foy}},
  \bibinfo {author} {\bibfnamefont {B.}~\bibnamefont {McCoy}}, \bibinfo
  {author} {\bibfnamefont {D.}~\bibnamefont {Holeman}}, \bibinfo {author}
  {\bibfnamefont {R.}~\bibnamefont {Mahoney}}, \bibinfo {author} {\bibfnamefont
  {G.}~\bibnamefont {Myers}}, \bibinfo {author} {\bibfnamefont
  {J.}~\bibnamefont {Herson}},\ and\ \bibinfo {author} {\bibfnamefont
  {T.}~\bibnamefont {Low}},\ }\bibfield  {title} {\bibinfo {title}
  {{Diamagnetically levitated robots: An approach to massively parallel robotic
  systems with unusual motion properties}},\ }\href
  {https://doi.org/10.1109/ICRA.2012.6225089} {\bibfield  {journal} {\bibinfo
  {journal} {Proceedings - IEEE International Conference on Robotics and
  Automation}\ ,\ \bibinfo {pages} {739}} (\bibinfo {year} {2012})}\BibitemShut
  {NoStop}%
\bibitem [{\citenamefont {Pelrine}\ \emph {et~al.}(2016)\citenamefont
  {Pelrine}, \citenamefont {Hsu}, \citenamefont {Wong-Foy}, \citenamefont
  {McCoy},\ and\ \citenamefont {Cowan}}]{Pelrine2016OptimalPatterns}%
  \BibitemOpen
  \bibfield  {author} {\bibinfo {author} {\bibfnamefont {R.}~\bibnamefont
  {Pelrine}}, \bibinfo {author} {\bibfnamefont {A.}~\bibnamefont {Hsu}},
  \bibinfo {author} {\bibfnamefont {A.}~\bibnamefont {Wong-Foy}}, \bibinfo
  {author} {\bibfnamefont {B.}~\bibnamefont {McCoy}},\ and\ \bibinfo {author}
  {\bibfnamefont {C.}~\bibnamefont {Cowan}},\ }\bibfield  {title} {\bibinfo
  {title} {{Optimal control of diamagnetically levitated milli robots using
  automated search patterns}},\ }in\ \href
  {https://doi.org/10.1109/MARSS.2016.7561725} {\emph {\bibinfo {booktitle}
  {2016 International Conference on Manipulation, Automation and Robotics at
  Small Scales (MARSS)}}}\ (\bibinfo  {publisher} {IEEE},\ \bibinfo {year}
  {2016})\ pp.\ \bibinfo {pages} {1--6}\BibitemShut {NoStop}%
\bibitem [{\citenamefont {Hsu}\ \emph {et~al.}(2017)\citenamefont {Hsu},
  \citenamefont {Cowan}, \citenamefont {Chu}, \citenamefont {McCoy},
  \citenamefont {Wong-Foy}, \citenamefont {Pelrine}, \citenamefont {Velez},
  \citenamefont {Arnold}, \citenamefont {Lake}, \citenamefont {Ballard},\ and\
  \citenamefont {Randall}}]{Hsu2017AutomatedMilli-robots}%
  \BibitemOpen
  \bibfield  {author} {\bibinfo {author} {\bibfnamefont {A.}~\bibnamefont
  {Hsu}}, \bibinfo {author} {\bibfnamefont {C.}~\bibnamefont {Cowan}}, \bibinfo
  {author} {\bibfnamefont {W.}~\bibnamefont {Chu}}, \bibinfo {author}
  {\bibfnamefont {B.}~\bibnamefont {McCoy}}, \bibinfo {author} {\bibfnamefont
  {A.}~\bibnamefont {Wong-Foy}}, \bibinfo {author} {\bibfnamefont
  {R.}~\bibnamefont {Pelrine}}, \bibinfo {author} {\bibfnamefont
  {C.}~\bibnamefont {Velez}}, \bibinfo {author} {\bibfnamefont
  {D.}~\bibnamefont {Arnold}}, \bibinfo {author} {\bibfnamefont
  {J.}~\bibnamefont {Lake}}, \bibinfo {author} {\bibfnamefont {J.}~\bibnamefont
  {Ballard}},\ and\ \bibinfo {author} {\bibfnamefont {J.}~\bibnamefont
  {Randall}},\ }\bibfield  {title} {\bibinfo {title} {{Automated 2D
  micro-assembly using diamagnetically levitated milli-robots}},\ }in\ \href
  {https://doi.org/10.1109/MARSS.2017.8001926} {\emph {\bibinfo {booktitle}
  {2017 International Conference on Manipulation, Automation and Robotics at
  Small Scales (MARSS)}}}\ (\bibinfo  {publisher} {IEEE},\ \bibinfo {year}
  {2017})\ pp.\ \bibinfo {pages} {1--6}\BibitemShut {NoStop}%
\bibitem [{\citenamefont {Pelrine}\ \emph {et~al.}(2019)\citenamefont
  {Pelrine}, \citenamefont {Hsu},\ and\ \citenamefont
  {Wong-Foy}}]{Pelrine2019MethodsDesigns}%
  \BibitemOpen
  \bibfield  {author} {\bibinfo {author} {\bibfnamefont {R.}~\bibnamefont
  {Pelrine}}, \bibinfo {author} {\bibfnamefont {A.}~\bibnamefont {Hsu}},\ and\
  \bibinfo {author} {\bibfnamefont {A.}~\bibnamefont {Wong-Foy}},\ }\bibfield
  {title} {\bibinfo {title} {{Methods and Results for Rotation of Diamagnetic
  Robots Using Translational Designs}},\ }in\ \href
  {https://doi.org/10.1109/MARSS.2019.8860975} {\emph {\bibinfo {booktitle}
  {2019 International Conference on Manipulation, Automation and Robotics at
  Small Scales (MARSS)}}}\ (\bibinfo  {publisher} {IEEE},\ \bibinfo {year}
  {2019})\ pp.\ \bibinfo {pages} {1--6}\BibitemShut {NoStop}%
\bibitem [{\citenamefont {Kuthan}\ \emph {et~al.}(2020)\citenamefont {Kuthan},
  \citenamefont {Jurik}, \citenamefont {Vitek},\ and\ \citenamefont
  {Mach}}]{Kuthan2020CollectiveOperation}%
  \BibitemOpen
  \bibfield  {author} {\bibinfo {author} {\bibfnamefont {J.}~\bibnamefont
  {Kuthan}}, \bibinfo {author} {\bibfnamefont {M.}~\bibnamefont {Jurik}},
  \bibinfo {author} {\bibfnamefont {M.}~\bibnamefont {Vitek}},\ and\ \bibinfo
  {author} {\bibfnamefont {F.}~\bibnamefont {Mach}},\ }\bibfield  {title}
  {\bibinfo {title} {{Collective Planar Actuation of Miniature Magnetic Robots
  Towards Individual Robot Operation}},\ }in\ \href
  {https://doi.org/10.1109/MARSS49294.2020.9307877} {\emph {\bibinfo
  {booktitle} {2020 International Conference on Manipulation, Automation and
  Robotics at Small Scales (MARSS)}}}\ (\bibinfo  {publisher} {IEEE},\ \bibinfo
  {year} {2020})\ pp.\ \bibinfo {pages} {1--6}\BibitemShut {NoStop}%
\bibitem [{\citenamefont {Kobayashi}\ and\ \citenamefont
  {Abe}(2012)}]{Kobayashi2012OpticalGraphite}%
  \BibitemOpen
  \bibfield  {author} {\bibinfo {author} {\bibfnamefont {M.}~\bibnamefont
  {Kobayashi}}\ and\ \bibinfo {author} {\bibfnamefont {J.}~\bibnamefont
  {Abe}},\ }\bibfield  {title} {\bibinfo {title} {{Optical motion control of
  maglev graphite}},\ }\href {https://doi.org/10.1021/ja310365k} {\bibfield
  {journal} {\bibinfo  {journal} {Journal of the American Chemical Society}\
  }\textbf {\bibinfo {volume} {134}},\ \bibinfo {pages} {20593} (\bibinfo
  {year} {2012})}\BibitemShut {NoStop}%
\bibitem [{\citenamefont {Han}\ \emph {et~al.}(2018)\citenamefont {Han},
  \citenamefont {Zhang}, \citenamefont {Chen},\ and\ \citenamefont
  {Sun}}]{Han2018Carbon-BasedActuators}%
  \BibitemOpen
  \bibfield  {author} {\bibinfo {author} {\bibfnamefont {B.}~\bibnamefont
  {Han}}, \bibinfo {author} {\bibfnamefont {Y.-L.}\ \bibnamefont {Zhang}},
  \bibinfo {author} {\bibfnamefont {Q.-D.}\ \bibnamefont {Chen}},\ and\
  \bibinfo {author} {\bibfnamefont {H.-B.}\ \bibnamefont {Sun}},\ }\bibfield
  {title} {\bibinfo {title} {{Carbon-Based Photothermal Actuators}},\ }\href
  {https://doi.org/10.1002/adfm.201802235} {\bibfield  {journal} {\bibinfo
  {journal} {Advanced Functional Materials}\ }\textbf {\bibinfo {volume}
  {28}},\ \bibinfo {pages} {1802235} (\bibinfo {year} {2018})}\BibitemShut
  {NoStop}%
\bibitem [{\citenamefont {Yang}\ \emph {et~al.}(2019)\citenamefont {Yang},
  \citenamefont {Yuan}, \citenamefont {Liu}, \citenamefont {Fang},
  \citenamefont {Fang}, \citenamefont {Yu},\ and\ \citenamefont
  {Li}}]{Yang2019PhotoresponsiveMaterials}%
  \BibitemOpen
  \bibfield  {author} {\bibinfo {author} {\bibfnamefont {M.}~\bibnamefont
  {Yang}}, \bibinfo {author} {\bibfnamefont {Z.}~\bibnamefont {Yuan}}, \bibinfo
  {author} {\bibfnamefont {J.}~\bibnamefont {Liu}}, \bibinfo {author}
  {\bibfnamefont {Z.}~\bibnamefont {Fang}}, \bibinfo {author} {\bibfnamefont
  {L.}~\bibnamefont {Fang}}, \bibinfo {author} {\bibfnamefont {D.}~\bibnamefont
  {Yu}},\ and\ \bibinfo {author} {\bibfnamefont {Q.}~\bibnamefont {Li}},\
  }\bibfield  {title} {\bibinfo {title} {{Photoresponsive Actuators Built from
  Carbon-Based Soft Materials}},\ }\href
  {https://doi.org/10.1002/adom.201900069} {\bibfield  {journal} {\bibinfo
  {journal} {Advanced Optical Materials}\ }\textbf {\bibinfo {volume} {7}},\
  \bibinfo {pages} {1900069} (\bibinfo {year} {2019})}\BibitemShut {NoStop}%
\bibitem [{\citenamefont {Young}\ \emph {et~al.}(2019)\citenamefont {Young},
  \citenamefont {Biggs}, \citenamefont {Yee},\ and\ \citenamefont
  {Elbidweihy}}]{Young2019OpticalGraphite}%
  \BibitemOpen
  \bibfield  {author} {\bibinfo {author} {\bibfnamefont {J.}~\bibnamefont
  {Young}}, \bibinfo {author} {\bibfnamefont {H.}~\bibnamefont {Biggs}},
  \bibinfo {author} {\bibfnamefont {S.}~\bibnamefont {Yee}},\ and\ \bibinfo
  {author} {\bibfnamefont {H.}~\bibnamefont {Elbidweihy}},\ }\bibfield  {title}
  {\bibinfo {title} {{Optical control and manipulation of diamagnetically
  levitated pyrolytic graphite}},\ }\bibfield  {journal} {\bibinfo  {journal}
  {AIP Advances}\ }\textbf {\bibinfo {volume} {9}},\ \href
  {https://doi.org/10.1063/1.5129886} {10.1063/1.5129886} (\bibinfo {year}
  {2019})\BibitemShut {NoStop}%
\bibitem [{\citenamefont {Ewall-Wice}\ \emph {et~al.}(2019)\citenamefont
  {Ewall-Wice}, \citenamefont {Yee}, \citenamefont {DeLawder}, \citenamefont
  {Montgomery}, \citenamefont {Joyce}, \citenamefont {Brownell},\ and\
  \citenamefont {ElBidweihy}}]{Ewall-Wice2019OptomechanicalGraphite}%
  \BibitemOpen
  \bibfield  {author} {\bibinfo {author} {\bibfnamefont {M.}~\bibnamefont
  {Ewall-Wice}}, \bibinfo {author} {\bibfnamefont {S.}~\bibnamefont {Yee}},
  \bibinfo {author} {\bibfnamefont {K.}~\bibnamefont {DeLawder}}, \bibinfo
  {author} {\bibfnamefont {S.~R.}\ \bibnamefont {Montgomery}}, \bibinfo
  {author} {\bibfnamefont {P.~J.}\ \bibnamefont {Joyce}}, \bibinfo {author}
  {\bibfnamefont {C.}~\bibnamefont {Brownell}},\ and\ \bibinfo {author}
  {\bibfnamefont {H.}~\bibnamefont {ElBidweihy}},\ }\bibfield  {title}
  {\bibinfo {title} {{Optomechanical Actuation of Diamagnetically Levitated
  Pyrolytic Graphite}},\ }\href {https://doi.org/10.1109/TMAG.2019.2892332}
  {\bibfield  {journal} {\bibinfo  {journal} {IEEE Transactions on Magnetics}\
  }\textbf {\bibinfo {volume} {55}},\ \bibinfo {pages} {1} (\bibinfo {year}
  {2019})}\BibitemShut {NoStop}%
\bibitem [{\citenamefont {Kim}\ \emph {et~al.}(2020)\citenamefont {Kim},
  \citenamefont {Pyo},\ and\ \citenamefont
  {Kim}}]{Kim2020HighlyMultistimuliResponses}%
  \BibitemOpen
  \bibfield  {author} {\bibinfo {author} {\bibfnamefont {J.~H.}\ \bibnamefont
  {Kim}}, \bibinfo {author} {\bibfnamefont {J.}~\bibnamefont {Pyo}},\ and\
  \bibinfo {author} {\bibfnamefont {T.}~\bibnamefont {Kim}},\ }\bibfield
  {title} {\bibinfo {title} {{Highly Mobile Levitating Soft Actuator Driven by
  Multistimuli-Responses}},\ }\href {https://doi.org/10.1002/admi.202001051}
  {\bibfield  {journal} {\bibinfo  {journal} {Advanced Materials Interfaces}\
  }\textbf {\bibinfo {volume} {7}},\ \bibinfo {pages} {2001051} (\bibinfo
  {year} {2020})}\BibitemShut {NoStop}%
\bibitem [{\citenamefont {Huang}\ \emph {et~al.}(2021)\citenamefont {Huang},
  \citenamefont {Yu}, \citenamefont {Su}, \citenamefont {Jiang}, \citenamefont
  {Chen},\ and\ \citenamefont {Shao}}]{Huang2021Light-responsiveApplications}%
  \BibitemOpen
  \bibfield  {author} {\bibinfo {author} {\bibfnamefont {Y.}~\bibnamefont
  {Huang}}, \bibinfo {author} {\bibfnamefont {Q.}~\bibnamefont {Yu}}, \bibinfo
  {author} {\bibfnamefont {C.}~\bibnamefont {Su}}, \bibinfo {author}
  {\bibfnamefont {J.}~\bibnamefont {Jiang}}, \bibinfo {author} {\bibfnamefont
  {N.}~\bibnamefont {Chen}},\ and\ \bibinfo {author} {\bibfnamefont
  {H.}~\bibnamefont {Shao}},\ }\bibfield  {title} {\bibinfo {title}
  {{Light-Responsive Soft Actuators: Mechanism, Materials, Fabrication, and
  Applications}},\ }\href {https://doi.org/10.3390/act10110298} {\bibfield
  {journal} {\bibinfo  {journal} {Actuators}\ }\textbf {\bibinfo {volume}
  {10}},\ \bibinfo {pages} {298} (\bibinfo {year} {2021})}\BibitemShut
  {NoStop}%
\bibitem [{\citenamefont {Yee}\ \emph {et~al.}(2021)\citenamefont {Yee},
  \citenamefont {Oney}, \citenamefont {Cosby}, \citenamefont {Durkin},\ and\
  \citenamefont {ElBidweihy}}]{Yee2021PhotothermalRevised}%
  \BibitemOpen
  \bibfield  {author} {\bibinfo {author} {\bibfnamefont {S.}~\bibnamefont
  {Yee}}, \bibinfo {author} {\bibfnamefont {L.}~\bibnamefont {Oney}}, \bibinfo
  {author} {\bibfnamefont {T.}~\bibnamefont {Cosby}}, \bibinfo {author}
  {\bibfnamefont {D.~P.}\ \bibnamefont {Durkin}},\ and\ \bibinfo {author}
  {\bibfnamefont {H.}~\bibnamefont {ElBidweihy}},\ }\bibfield  {title}
  {\bibinfo {title} {{Photothermal actuation of levitated pyrolytic graphite
  revised}},\ }\href {https://doi.org/10.1063/5.0061390} {\bibfield  {journal}
  {\bibinfo  {journal} {APL Materials}\ }\textbf {\bibinfo {volume} {9}},\
  \bibinfo {pages} {101107} (\bibinfo {year} {2021})}\BibitemShut {NoStop}%
\bibitem [{\citenamefont {Shen}\ \emph {et~al.}(2022)\citenamefont {Shen},
  \citenamefont {Wu}, \citenamefont {Yang}, \citenamefont {Yang}, \citenamefont
  {Liu},\ and\ \citenamefont {Wu}}]{Shen2022OpticalGraphite}%
  \BibitemOpen
  \bibfield  {author} {\bibinfo {author} {\bibfnamefont {S.}~\bibnamefont
  {Shen}}, \bibinfo {author} {\bibfnamefont {L.}~\bibnamefont {Wu}}, \bibinfo
  {author} {\bibfnamefont {S.}~\bibnamefont {Yang}}, \bibinfo {author}
  {\bibfnamefont {Q.}~\bibnamefont {Yang}}, \bibinfo {author} {\bibfnamefont
  {J.-T.}\ \bibnamefont {Liu}},\ and\ \bibinfo {author} {\bibfnamefont
  {Z.}~\bibnamefont {Wu}},\ }\bibfield  {title} {\bibinfo {title} {{Optical
  energy harvesting in vibrate maglev graphite}},\ }\href
  {https://doi.org/10.1016/j.carbon.2021.11.016} {\bibfield  {journal}
  {\bibinfo  {journal} {Carbon}\ }\textbf {\bibinfo {volume} {187}},\ \bibinfo
  {pages} {266} (\bibinfo {year} {2022})}\BibitemShut {NoStop}%
\bibitem [{\citenamefont {Vikrant}\ and\ \citenamefont
  {Jayanth}(2022)}]{Vikrant2022DiamagneticallyFreedom}%
  \BibitemOpen
  \bibfield  {author} {\bibinfo {author} {\bibfnamefont {K.~S.}\ \bibnamefont
  {Vikrant}}\ and\ \bibinfo {author} {\bibfnamefont {G.~R.}\ \bibnamefont
  {Jayanth}},\ }\bibfield  {title} {\bibinfo {title} {{Diamagnetically
  levitated nanopositioners with large-range and multiple degrees of
  freedom}},\ }\href {https://doi.org/10.1038/s41467-022-31046-4} {\bibfield
  {journal} {\bibinfo  {journal} {Nature Communications}\ }\textbf {\bibinfo
  {volume} {13}},\ \bibinfo {pages} {3334} (\bibinfo {year}
  {2022})}\BibitemShut {NoStop}%
\bibitem [{\citenamefont {Lewandowski}\ \emph
  {et~al.}(2021{\natexlab{b}})\citenamefont {Lewandowski}, \citenamefont
  {Knowles}, \citenamefont {Etienne},\ and\ \citenamefont
  {D'Urso}}]{Lewandowski2021High-SensitivityMicrosphere}%
  \BibitemOpen
  \bibfield  {author} {\bibinfo {author} {\bibfnamefont {C.~W.}\ \bibnamefont
  {Lewandowski}}, \bibinfo {author} {\bibfnamefont {T.~D.}\ \bibnamefont
  {Knowles}}, \bibinfo {author} {\bibfnamefont {Z.~B.}\ \bibnamefont
  {Etienne}},\ and\ \bibinfo {author} {\bibfnamefont {B.}~\bibnamefont
  {D'Urso}},\ }\bibfield  {title} {\bibinfo {title} {{High-Sensitivity
  Accelerometry with a Feedback-Cooled Magnetically Levitated Microsphere}},\
  }\bibfield  {journal} {\bibinfo  {journal} {Physical Review Applied}\
  }\textbf {\bibinfo {volume} {15}},\ \href
  {https://doi.org/10.1103/PhysRevApplied.15.014050}
  {10.1103/PhysRevApplied.15.014050} (\bibinfo {year}
  {2021}{\natexlab{b}})\BibitemShut {NoStop}%
\bibitem [{\citenamefont {Nakashima}(2020)}]{Nakashima2020}%
  \BibitemOpen
  \bibfield  {author} {\bibinfo {author} {\bibfnamefont {R.}~\bibnamefont
  {Nakashima}},\ }\bibfield  {title} {\bibinfo {title} {{Diamagnetic levitation
  of a milligram-scale silica using permanent magnets for the use in a
  macroscopic quantum measurement}},\ }\href
  {https://doi.org/10.1016/j.physleta.2020.126592} {\bibfield  {journal}
  {\bibinfo  {journal} {Physics Letters A}\ }\textbf {\bibinfo {volume}
  {384}},\ \bibinfo {pages} {126592} (\bibinfo {year} {2020})}\BibitemShut
  {NoStop}%
\bibitem [{\citenamefont {Leng}\ \emph {et~al.}(2021)\citenamefont {Leng},
  \citenamefont {Li}, \citenamefont {Kong}, \citenamefont {Xie}, \citenamefont
  {Zheng}, \citenamefont {Yin}, \citenamefont {Xiong}, \citenamefont {Wu},
  \citenamefont {Duan}, \citenamefont {Du}, \citenamefont {Yin}, \citenamefont
  {Huang},\ and\ \citenamefont {Du}}]{Leng2021}%
  \BibitemOpen
  \bibfield  {author} {\bibinfo {author} {\bibfnamefont {Y.}~\bibnamefont
  {Leng}}, \bibinfo {author} {\bibfnamefont {R.}~\bibnamefont {Li}}, \bibinfo
  {author} {\bibfnamefont {X.}~\bibnamefont {Kong}}, \bibinfo {author}
  {\bibfnamefont {H.}~\bibnamefont {Xie}}, \bibinfo {author} {\bibfnamefont
  {D.}~\bibnamefont {Zheng}}, \bibinfo {author} {\bibfnamefont
  {P.}~\bibnamefont {Yin}}, \bibinfo {author} {\bibfnamefont {F.}~\bibnamefont
  {Xiong}}, \bibinfo {author} {\bibfnamefont {T.}~\bibnamefont {Wu}}, \bibinfo
  {author} {\bibfnamefont {C.-K.}\ \bibnamefont {Duan}}, \bibinfo {author}
  {\bibfnamefont {Y.}~\bibnamefont {Du}}, \bibinfo {author} {\bibfnamefont
  {Z.-q.}\ \bibnamefont {Yin}}, \bibinfo {author} {\bibfnamefont
  {P.}~\bibnamefont {Huang}},\ and\ \bibinfo {author} {\bibfnamefont
  {J.}~\bibnamefont {Du}},\ }\bibfield  {title} {\bibinfo {title} {{Mechanical
  Dissipation Below 1 {$\mu$}Hz with a Cryogenic Diamagnetic Levitated
  Micro-Oscillator}},\ }\href
  {https://doi.org/10.1103/PhysRevApplied.15.024061} {\bibfield  {journal}
  {\bibinfo  {journal} {Physical Review Applied}\ }\textbf {\bibinfo {volume}
  {15}},\ \bibinfo {pages} {024061} (\bibinfo {year} {2021})}\BibitemShut
  {NoStop}%
\bibitem [{\citenamefont {Niu}\ \emph {et~al.}(2018)\citenamefont {Niu},
  \citenamefont {Lin}, \citenamefont {Wang}, \citenamefont {Bao},\ and\
  \citenamefont {Hu}}]{Niu2018}%
  \BibitemOpen
  \bibfield  {author} {\bibinfo {author} {\bibfnamefont {C.}~\bibnamefont
  {Niu}}, \bibinfo {author} {\bibfnamefont {F.}~\bibnamefont {Lin}}, \bibinfo
  {author} {\bibfnamefont {Z.~M.}\ \bibnamefont {Wang}}, \bibinfo {author}
  {\bibfnamefont {J.}~\bibnamefont {Bao}},\ and\ \bibinfo {author}
  {\bibfnamefont {J.}~\bibnamefont {Hu}},\ }\bibfield  {title} {\bibinfo
  {title} {{Graphene levitation and orientation control using a magnetic
  field}},\ }\href {https://doi.org/10.1063/1.5005539} {\bibfield  {journal}
  {\bibinfo  {journal} {Journal of Applied Physics}\ }\textbf {\bibinfo
  {volume} {123}},\ \bibinfo {pages} {044302} (\bibinfo {year}
  {2018})}\BibitemShut {NoStop}%
\bibitem [{\citenamefont {Chen}\ \emph {et~al.}(2022)\citenamefont {Chen},
  \citenamefont {Ammu}, \citenamefont {Masania}, \citenamefont {Steeneken},\
  and\ \citenamefont {Alijani}}]{Chen2022DiamagneticResonators}%
  \BibitemOpen
  \bibfield  {author} {\bibinfo {author} {\bibfnamefont {X.}~\bibnamefont
  {Chen}}, \bibinfo {author} {\bibfnamefont {S.~K.}\ \bibnamefont {Ammu}},
  \bibinfo {author} {\bibfnamefont {K.}~\bibnamefont {Masania}}, \bibinfo
  {author} {\bibfnamefont {P.~G.}\ \bibnamefont {Steeneken}},\ and\ \bibinfo
  {author} {\bibfnamefont {F.}~\bibnamefont {Alijani}},\ }\bibfield  {title}
  {\bibinfo {title} {{Diamagnetic Composites for High-Q Levitating
  Resonators}},\ }\href {https://doi.org/10.1002/advs.202203619} {\bibfield
  {journal} {\bibinfo  {journal} {Advanced Science}\ }\textbf {\bibinfo
  {volume} {2203619}},\ \bibinfo {pages} {2203619} (\bibinfo {year}
  {2022})}\BibitemShut {NoStop}%
\bibitem [{\citenamefont {Wang}\ and\ \citenamefont
  {Uhlenbeck}(1945)}]{Wang1945OnII}%
  \BibitemOpen
  \bibfield  {author} {\bibinfo {author} {\bibfnamefont {M.~C.}\ \bibnamefont
  {Wang}}\ and\ \bibinfo {author} {\bibfnamefont {G.~E.}\ \bibnamefont
  {Uhlenbeck}},\ }\bibfield  {title} {\bibinfo {title} {{On the theory of the
  Brownian motion II}},\ }\href@noop {} {\bibfield  {journal} {\bibinfo
  {journal} {Reviews of modern physics}\ }\textbf {\bibinfo {volume} {17}},\
  \bibinfo {pages} {323} (\bibinfo {year} {1945})}\BibitemShut {NoStop}%
\bibitem [{\citenamefont {Kirpo}\ \emph {et~al.}(2010)\citenamefont {Kirpo},
  \citenamefont {Boeck},\ and\ \citenamefont
  {Thess}}]{Kirpo2010Eddy-currentDipole}%
  \BibitemOpen
  \bibfield  {author} {\bibinfo {author} {\bibfnamefont {M.}~\bibnamefont
  {Kirpo}}, \bibinfo {author} {\bibfnamefont {T.}~\bibnamefont {Boeck}},\ and\
  \bibinfo {author} {\bibfnamefont {A.}~\bibnamefont {Thess}},\ }\bibfield
  {title} {\bibinfo {title} {{Eddy-current braking of a translating solid bar
  by a magnetic dipole}},\ }\href {https://doi.org/10.1002/pamm.201010249}
  {\bibfield  {journal} {\bibinfo  {journal} {Pamm}\ }\textbf {\bibinfo
  {volume} {10}},\ \bibinfo {pages} {513} (\bibinfo {year} {2010})}\BibitemShut
  {NoStop}%
\bibitem [{\citenamefont {Votyakov}\ and\ \citenamefont
  {Thess}(2012)}]{Votyakov2012InteractionPlate}%
  \BibitemOpen
  \bibfield  {author} {\bibinfo {author} {\bibfnamefont {E.~V.}\ \bibnamefont
  {Votyakov}}\ and\ \bibinfo {author} {\bibfnamefont {A.}~\bibnamefont
  {Thess}},\ }\bibfield  {title} {\bibinfo {title} {{Interaction of a magnetic
  dipole with a slowly moving electrically conducting plate}},\ }\href
  {https://doi.org/10.1007/s10665-012-9545-1} {\bibfield  {journal} {\bibinfo
  {journal} {Journal of Engineering Mathematics}\ }\textbf {\bibinfo {volume}
  {77}},\ \bibinfo {pages} {147} (\bibinfo {year} {2012})}\BibitemShut
  {NoStop}%
\bibitem [{\citenamefont {Carlstedt}\ \emph {et~al.}(2014)\citenamefont
  {Carlstedt}, \citenamefont {Porzig}, \citenamefont {Ziolkowski},
  \citenamefont {Uhlig}, \citenamefont {Brauer},\ and\ \citenamefont
  {Toepfer}}]{Carlstedt2014ComparisonSimulations}%
  \BibitemOpen
  \bibfield  {author} {\bibinfo {author} {\bibfnamefont {M.}~\bibnamefont
  {Carlstedt}}, \bibinfo {author} {\bibfnamefont {K.}~\bibnamefont {Porzig}},
  \bibinfo {author} {\bibfnamefont {M.}~\bibnamefont {Ziolkowski}}, \bibinfo
  {author} {\bibfnamefont {R.~P.}\ \bibnamefont {Uhlig}}, \bibinfo {author}
  {\bibfnamefont {H.}~\bibnamefont {Brauer}},\ and\ \bibinfo {author}
  {\bibfnamefont {H.}~\bibnamefont {Toepfer}},\ }\bibfield  {title} {\bibinfo
  {title} {{Comparison of Lorentz force eddy current testing and common eddy
  current testing-measurements and simulations}},\ }\href
  {https://doi.org/10.3233/978-1-61499-407-7-218} {\bibfield  {journal}
  {\bibinfo  {journal} {Studies in Applied Electromagnetics and Mechanics}\
  }\textbf {\bibinfo {volume} {39}},\ \bibinfo {pages} {218} (\bibinfo {year}
  {2014})}\BibitemShut {NoStop}%
\end{thebibliography}
%

\end{document}


\preprint{AIP/123-QED}
\title{Supplementary Material}
\author{P. Romagnoli}
\thanks{These three authors contributed equally}
\affiliation{ 
Quantum Machines Unit, Okinawa Institute of Science and Technology Graduate University, Onna, Okinawa 904-0495, Japan \looseness=-1
}%
\author{R. Lecamwasam}
\thanks{These three authors contributed equally}
\affiliation{ 
Quantum Machines Unit, Okinawa Institute of Science and Technology Graduate University, Onna, Okinawa 904-0495, Japan \looseness=-1
}
\author{S. Tian}
\thanks{These three authors contributed equally}
\affiliation{ 
Quantum Machines Unit, Okinawa Institute of Science and Technology Graduate University, Onna, Okinawa 904-0495, Japan \looseness=-1
}%
\author{J.E. Downes}
\affiliation{
School of Mathematical and Physical Science, Macquarie University, 2109 NSW, Australia \looseness=-1
}

\author{J. Twamley}%
\email{jason.twamley@oist.jp}
\affiliation{ 
Quantum Machines Unit, Okinawa Institute of Science and Technology Graduate University, Onna, Okinawa 904-0495, Japan \looseness=-1
}%
\maketitle

\section{Experimental setup} \label{SupplementaryExperiment}

The NdFeB magnets are grade N52 (B888-N52) from K\&J Magnetics. The magnets are rigidly held within a holder which is fixed to a five axis vacuum-compatible motorised stage. The magnet/motor platform is mounted on a small optical breadboard which in turn sits on four wire rope vibration isolation supports.

The graphite samples have a size of $10\:\,{\rm mm}\times10\:\,{\rm mm}\times\sim0.77\:\,{\rm mm}$. These were machined from a single piece of graphite (PG3, K\&J Magnetics), to ensure that all samples possessed similar electric and magnetic properties. Note that the thickness of the graphite is not uniform due to the very coarse surface. The ring-like slit patterns were machined by a femtosecond laser machining through the graphite samples. 

A small mirror is fixed to the centre of each machined graphite sample. These mirrors are fabricated by sputtering an aluminium coating onto a microscope coverglass (thickness $\sim0.17\:\,{\rm mm}$), and then cut in to pieces of dimension $3\:{\rm mm}\times3\:{\rm mm}$. The mirrors then are attached to the graphite plate by optical glue. Following machining and gluing of the mirror, the masses of the plates $N=(0,3,6,8)$ were $(147,137,93,79)\:{\rm mg}$ respectively.

A PICOSCALE laser interferometic displacement sensor is used to measure the vertical displacement of the sample relative to an optomechanical cage system arranged to hold the sensor head above the graphite sample. The cage structure is fixed to the optical breadboard. The laser is aligned using the 5-axis stage to reflect from the mirror which is affixed to the centre of the graphite sample. The displacement sensor is based on a compact Michelson interferometer, and enables high precision measurements in real time with a resolution of picometers at a high bandwidth. A graphical user interface of the interferometer is used to acquire position and velocity data, and data is streamed to a PC for storage and later analysis. The position is also output from the interferometer to an oscilloscope for monitoring.

The whole structure (magnets/motor/optical sensor/breadboard/vibration isolators) is positioned inside a vacuum chamber. The optical signal from the in-vacuum interferometer sensor head is fed via an optical fibre through a feedthrough to the external interferometer. The chamber is evacuated by a system consisting of a turbopump (HiCube300H, Pfeiffer), an ion pump (VacIon Plus 300, Agilent), and associated roughing pump. The vacuum chamber and ion pump are supported by a vibration isolated/damped optical table, while the turbopump is supported by a separate vibration isolation platform. The ion pump has no moving parts and is motionally extremely quiet, while the turbopump rests on its own vibration isolated support platform. The turbopump is used initially to reach a base pressure ($10^{-6}\:{\rm hPa}$), then we switch to the ion pump to reach high vacuum ($10^{-7}\:{\rm hPa}$). During measurement periods a gate-valve connected to the turbopump is closed and the turboppump is switched off to avoid unwanted mechanical vibrations. The ion-pump then operates continuously to maintain high vacuum. 

\section{Estimating motional frequencies for the $N=0$ plate}\label{Supplementarynormalmodes}
To understand the measured power spectral density, we simulated the oscillation frequencies of the $N=0$ plate. These modes are determined by the forces and torques the plate experiences in the inhomogenous magnetic field above the checkerboard magnet array. Due to the small size of the slits, the modes are not expected to be significantly different for other values of $N$. In the following, we will use "laboratory frame" to denote the reference frame of the magnets.

A diamagnetic material sitting in a magnetic field ${\bf B}$ experiences an effective induced magnetization ${\bf M}=(1/\mu_0){\bf \chi}\cdot {\bf B}$, where ${\bf \chi}$ is the magnetic susceptibility tensor for the material. In the body frame of the plate this tensor is diagonal: ${\bf \chi}={\rm diag}(\chi_x, \chi_y, \chi_z)$, with values $\chi_j$ given in  \Cref{BigTable}. The force density on the graphite plate is then\cite{Niu2018} ${\bf f}=\left({\bf M}\cdot\nabla\right){\bf B}$. If the plate's normal is vertical in the laboratory frame, then we have 
\begin{equation}
    {\bf f}=\frac{1}{2\mu_0}\nabla(\chi_x\,B_x^2+\chi_y\,B_y^2+\chi_z\,B_z^2)\;\;,
\end{equation}
but if the plate is tilted then the $\chi-$tensor, which is diagonal in the body frame of the plate, has to be transformed into the laboratory frame. 

Analytic forms for the magnetic fields arising from the checkerboard array are given in \S 2.5 of \citeinline{Camacho2013}, though the expressions are too complex to be reproduced here. We then numerically (in Matlab) evaluate the force density over a dense 3D grid throughout the graphite plate, which we integrate numerically to obtain the net force on the centre of mass of the plate as a function of the position and angular tilt. Setting the net force ${\bf F_{\rm net}}={\bf F}_{\rm mag}+{\bf F}_{\rm grav}=0$, we obtain the equilibrium position of the centre of mass of the plate as $\vec{r}=(0,0,z_{\rm equil})$ (see Table \ref{BigTable}). 

\begin{figure}
	\includegraphics[width=\linewidth]{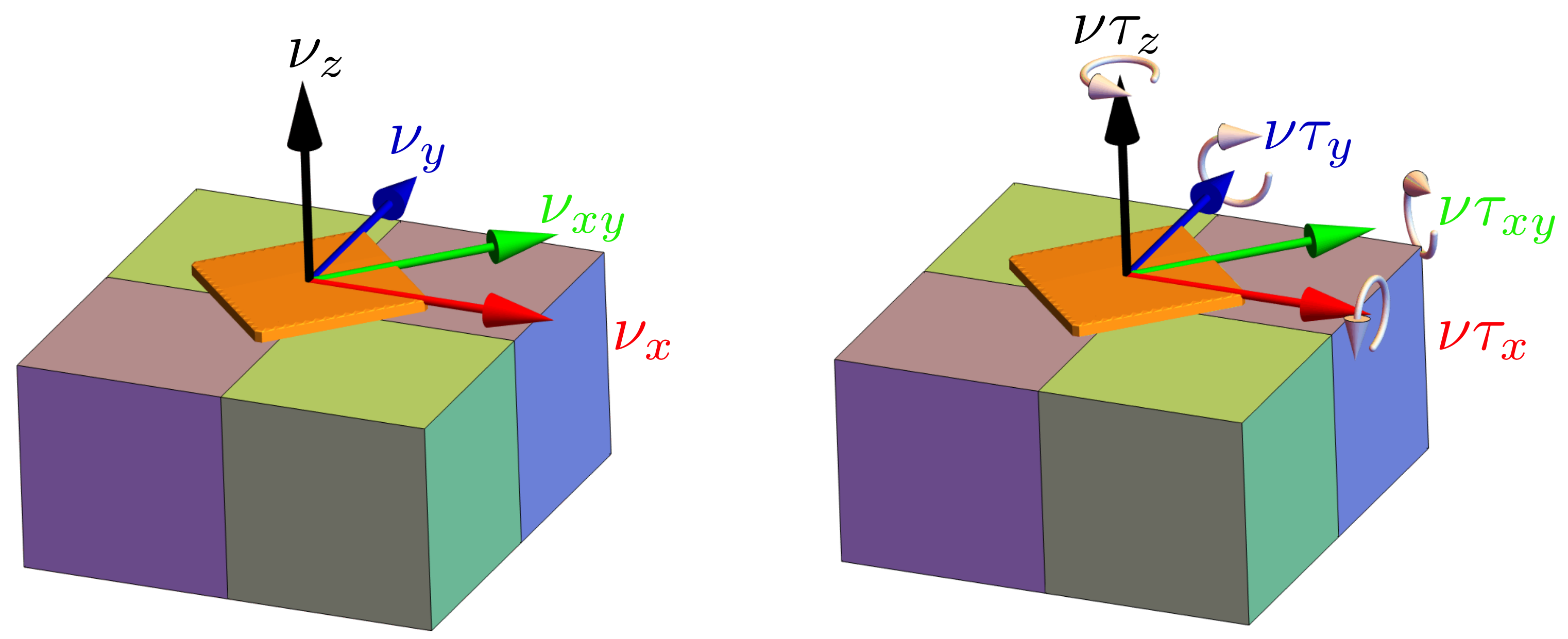}

\caption{Schematic of normal mode analysis: we illustrate the displacement (left) and torsional (right), motions of the levitated graphite plate (orange), above the 2x2 checkerboard magnet array (purple and green): (a) we show the displacement restoring frequencies along the \{ $x$, $y$, $z$, $x+y$\} directions (red, blue, black, green); (b) we show the torsional restoring frequencies for oscillations about the same axes as in (a). The normal mode motional frequencies for the solid graphite plate are reported in Table \ref{BigTable}.  } \label{fig:NormalModes}
\end{figure}

To evaluate the equilibrium orientation of the plate one can use the torque density exerted by the interaction of the induced magnetization of the plate with the magnetic field of the magnet array: 
\begin{equation}
    {\bf \tau}={\bf M}\times {\bf B}+{\bf r}\times {\bf f}\;\;,
\end{equation}
where one again must take care to transform the $\chi$-tensor into the laboratory frame. The total torque on the plate is obtained by integrating this torque density over the volume of the plate. We find that this vanishes and yields a stable equilibrium value for the configuration where the plate is horizontal but rotated by $\pi/4$, around the $z$-axis relative to the checkerboard frame, as shown in Fig \ref{fig:NormalModes}.

To determine the oscillation frequencies of the plate when subject to small displacements or tilts from its equilibrium configuration, we evaluated the net force or torque as a function of these small perturbations, then evaluated the spring stiffness via:
\begin{equation}
    {\bf \delta F_{\rm NET}}=-k{\bf \delta r},\; {\bf \delta \tau_{\rm NET}}=-k{\bf \delta \theta}\;\;.
\end{equation}
The oscillation frequencies where then given by
\begin{equation}
    2\pi \nu_j=\sqrt{\frac{k}{m}},\; 2\pi \nu\tau_j=\sqrt{\frac{k}{I_j}}\;\;,
\end{equation}
where $m$ is the total mass of the plate and $I_j$ is the moment of inertia of the plate about axis $j$. 

The various directions of displacements and axes for torsional oscillations studied are shown in Fig \ref{fig:NormalModes}, which can be compared with the numerical in Table \ref{BigTable}. We performed an experiment to confirm these numerical predictions, and observed visually the center of mass of the $N=8$ graphite oscillating in the vertical oscillation mode of motion with a frequency $\nu_z\sim 17\,{\rm Hz}$ (see Supplementary video).
 The simulations also showed evidence for a small coupling between the displacement and torsional modes.

\section{Modelling the power spectral density}\label{supp:ModellingPSD}
We approximate the plate's dynamics as a one-dimensional harmonic oscillator with some effective damping rate $\gamma$, whose natural frequency is approximately $f_0=17{\rm Hz}$. This has equation of motion (see \S 10 of \citeinline{Wang1945OnII})
\begin{equation}\label{eq:DampedOscillator}
    z''(t)+\frac{\gamma}{m}z'(t)+(2\pi f_0)^2z(t)=\sqrt{\frac{2\gamma k_BT}{m}}\xi(t)\;\;,
\end{equation}
where $m$ is the effective mass, $k_B$ Boltzmann's constant, and $T$ the effective temperature of the mode. The term $\xi(t)$ represents white noise with correlation $\langle\xi(t)\xi(t')\rangle=\delta(t-t')$, and describes a stochastic force due to thermal fluctuations in the electrons in the graphite plate, which are heated by the eddy currents. 

The power spectral density $S_{zz}$ is given by\cite{Wang1945OnII}
\begin{equation}\label{eq:BrownianMotionPSD}
    S_{zz}(f)=\frac{8k_BT\gamma/m}{((2\pi f_0)^2-(2\pi f)^2)^2+(2\pi f \gamma/m )^2}\;\;.
\end{equation}
This forms a shape similar to that of a Lorentzian \footnote{This distribution is often referred to as \emph{Lorentzian}, however it is slightly different since $\gamma$ on the denominator is multiplied by $f$.}, whose width increases with $\gamma/m$. The distribution is peaked at frequency
\begin{equation}
    f=\sqrt{f_0^2-\frac{\gamma^2}{8m^2\pi^2}}\;\;.
\end{equation}

We took position data for 1200 seconds with time interval $\Delta t=0.41\:{\rm ms}$, and performed a Welch Peridogram analysis over blocks of size $60\:{\rm s}$, which averages out noise in the PSD. This data was fitted to \cref{eq:BrownianMotionPSD}, to obtain estimates of the parameters $f_0$ and $\gamma$. The resulting PSDs and fittings are shown in the main text in Fig. 3 
We note that as the slot density increases the mechanical damping decreases, and there is a small shift in $f_0$ towards larger values. This shift is most likely due to the small change in trapping forces and sample mass due to the removal of the slot material and various variations in the sample thickness.

Another possible source of damping of the resonator is collisions with background gas molecules. To estimate the quality factor $Q_g$ associated with the gas damping, we will simplify the geometry of our sample and assume it is a graphite sphere of radius of $r=1\:{\rm mm}$ surrounded by nitrogen gas molecules ($m_{g}=4.65\times10^{-26}\: {\rm Kg}$), at a pressure of $p_g=1\times10^{-7}\:{\rm hPa}$. For a graphite sphere of density $\rho$ undergoing Brownian motion in a harmonic trap of frequency $\omega_z=2\pi \nu_z$, the quality factor is given by \cite{Wang2019}
\begin{equation}
Q_{g}=\frac{\pi\omega_z}{6}\frac{\rho r}{p_{g}}\sqrt{\frac{3k_{B}T}{m_{g}}}\;\;.\label{Q-gass}
\end{equation}
Using values for $\rho$ and $\omega_z$ from \Cref{BigTable}, we obtain $Q_g\sim 6\times10^9$. This is much larger than the quality factor expected from eddy damping. Thus we may treat air damping as negligible.

\section{Simulating the eddy currents}\label{Supplementary2Deddy}

\subsection{Two-dimensional Mathematica model}
The Mathematica simulation modelled the eddy currents by solving Poisson's equation for an effective electric potential, following the method in \citeinline{Chen2020}. As a conductor moves with velocity $\bf v$ through a magnetic field $\bf B$, the delocalised electrons feel a Lorentz force proportional to ${\bf v}\times {\bf B}$. This moves the electrons and causes charge to collect at the boundaries, resulting in an electric potential $V_e$. The eddy current density $\bf j$ is formed by the sum of these electric and magnetic forces:
\begin{equation}\label{eq:EddyCurrentDensity}
    {\bf j}=-{\bf \sigma}{\bf \nabla} V_e+{\bf \sigma} ({\bf v}\times {\bf B})\;\;,
\end{equation}
where $\sigma$ is the $3\times 3$ electrical conductivity matrix. Only the diagonal entries $\sigma_x,\sigma_y,\sigma_z$ are non-zero, describing how easily electrons in the material may flow in each Cartesian direction. On the right hand side, $\mathbf{B}$ (the external magnetic field) and $\mathbf{v}$ (the plate velocity) are known, so to find $\bf j$ we must solve for the electric potential $V_e$.

Maxwell's equations imply conservation of charge: ${\bf\nabla}\cdot {\bf j}=0$. Applying this to \cref{eq:EddyCurrentDensity}, we find that the potential satisfies Poisson's equation:
\begin{equation}\label{eq:EddyCurrentPDE}
   {\bf \nabla}\cdot\left[-{\bf \sigma\nabla} V_e+{\bf \sigma}({\bf v}\times {\bf B})\right]=0\;\;.
\end{equation}
The conductivity $\sigma$ does not cancel, since it is a diagonal matrix whose entries are not all equal. The boundary condition comes from the fact that current must flow parallel to the edges of the conductor. Thus if $\bf \hat{n}$ is the unit vector for the boundary, we have $\bf j\cdot \hat{n}=0$, which expressed in terms of the potential gives:
\begin{equation}\label{eq:EddyCurrentBoundaryCondition}
    {\bf \hat{n}}\cdot({\bf \sigma\nabla} V_e)={\bf \hat{n}}\cdot{\bf \sigma(v\times B)}\;\;.
\end{equation}
The partial differential equation (\ref{eq:EddyCurrentPDE}), with the Neumann boundary condition (\ref{eq:EddyCurrentBoundaryCondition}), has a solution $V_e$ which is unique up to addition of a constant. This constant does not affect the eddy currents, since $\bf j$ is proportional to the gradient ${\bf\nabla} V_e$.

For graphite, the conductivity in the $xy$-plane is much larger than that along the $z$-direction. We can thus define $\sigma_{xy}=\sigma_x=\sigma_y$, and approximate $\sigma_z\approx 0$. Then $\bf \sigma$ cancels from (\ref{eq:EddyCurrentPDE}), which becomes
\begin{equation}\label{eq:EddyCurrentPDE2D}
    (\partial_x^2+\partial_y^2)V_e(x,y)={\bf \nabla}\cdot ({\bf v}\times {\bf B})_{xy}\;\;,
\end{equation}
with boundary condition
\begin{equation}\label{eq:EddyCurrentBoundaryCondition2D}
    {\bf \hat{n}}\cdot{\bf\nabla} V_e(x,y)={\bf\hat{n}}\cdot({\bf v}\times {\bf B})_{xy}\;\;,
\end{equation}
where $({\bf v}\times {\bf B})_{xy}$ denotes the projection of the Lorentz force into the $xy$-plane. The resulting eddy current density is then
\begin{equation}
    {\bf j}(x,y)=-\sigma_{xy}\left[{\bf \nabla} V_e(x,y)+({\bf v}\times {\bf B})_{xy}\right]\;\;.
    \label{eq:2DEddyCurrent}
\end{equation}

Both quantities on the right-hand side of \cref{eq:EddyCurrentPDE2D} are known. The magnetic field is generated by the checkerboard of permanent magnets, evaluated at the $z$-value corresponding to the equilibrium position of the levitated plate. An analytical expression for the magnetic field of a rectangular prism can be found in \S10 of \citeinline{Camacho2013}, from which we can calculate $\bf B$ for the array. Since we are considering vertical motion, the velocity can be taken to be ${\bf v}=(0,0,1)$. Note that the magnitude of $\bf v$ doesn't matter, since from  \cref{eq:EddyCurrentPDE2D} and \cref{eq:EddyCurrentBoundaryCondition2D}  we can see that $V_e$ is linearly proportional to $\bf v$.

\cref{eq:EddyCurrentPDE2D} and \cref{eq:EddyCurrentBoundaryCondition2D} were solved using the Mathematica's finite element method functionality. The results were found to be stable for a grid size going from $1\mathrm{mm}$ to $0.05\mathrm{mm}$, indicating convergence of the numerical method. 

The current density $\bf j$ can then be used to calculate the eddy damping \cite{Kirpo2010Eddy-currentDipole, Votyakov2012InteractionPlate, Carlstedt2014ComparisonSimulations, Chen2020}, and hence the estimated change in quality factor. The electrons in the current will feel a Lorentz force $\mathbf{j}\times\mathbf{B}$ due to the permanent magnets below. Integrating this over the entire plate gives us the net force due to the eddy currents:
\begin{equation}
    {\bf F}_{\rm Eddy}=\int({\bf j}\times {\bf B})\,{\rm d}S\;\;,
\end{equation}
where $\int {\rm d}S$ represents an integral over the two-dimensional plate. We find the net eddy force to be directed downwards, opposing the upwards motion $\bf v$. Since $V_e$ is linearly proportional to $\bf v$, so is $\bf j$, and hence $\bf F_{\rm eddy}$. Thus we may write
\begin{equation}
    {\bf F}_{\rm eddy}=-\gamma_{\rm eddy}{\bf v}\;\;,
\end{equation}
where $\gamma_{\rm eddy}$ is a positive constant associated with each plate geometry, which we recognise as the damping on the plate due to eddy currents.

\subsection{Three-dimensional COMSOL model}

A three-dimensional COMSOL model was constructed using the parameters in \Cref{BigTable}. The static magnetic force on the graphite plate was calculated within COMSOL using the force equations described in  \cref{Supplementarynormalmodes}. The equilibrium levitation height determined by balancing the magnetic force with gravity agrees well with both the  calculations performed in \cref{Supplementarynormalmodes}, and the experiment. Then,  using a time-dependent study in COMSOL, we simulated the eddy current damping by calculating the Lorentz force at the equilibrium position as a function of velocity. We show the ratio of these COMSOL derived $Q-$factors in the main text Fig. 4. 
while examples of the current distributions in a 2D horizontal slice are shown in the main text Fig. 5. 

\begin{table*}[hbt!]
\begin{center}
\caption{Quantities and parameter values appearing in this manuscript - bold items are the results of simulation while non-bold symbols are measured or taken from the literature.\label{BigTable}}
\begin{threeparttable}
\begin{tabular}{ p{10cm}p{2.5cm}p{3cm}p{1cm}}
\hline\hline
 Quantity & Symbol &   Value & Unit   \\ \hline
{\bf Displacement} frequency along $x$: & {\bf $\bf\nu_x$} &   3.9 & {\rm Hz}    \\
{\bf Displacement} frequency along $y$: & {\bf $\bf\nu_y$} &   3.9  & {\rm Hz}     \\
{\bf Displacement} frequency along $z$: & {\bf $\bf\nu_z$} &   17.2 & {\rm Hz}     \\
{\bf Displacement} frequency along $x+y$: & {\bf $\bf\nu_{xy}$} &   4.1 & {\rm Hz}   \\ \hline
{\bf Torsional} frequency about $x$: & {\bf $\bf\nu\tau_x$} &   17.5  & {\rm Hz}   \\
{\bf Torsional} frequency about $y$: & {\bf $\bf\nu\tau_y$} &    17.5 &  {\rm Hz}  \\
{\bf Torsional} frequency about $z$: & {\bf $\bf\nu\tau_z$} &   5.9 &  {\rm Hz}  \\
{\bf Torsional} frequency about $x+y$: & {\bf $\bf\nu\tau_{xy}$} &   16.9 &  {\rm Hz}  \\\hline
Magnetic Susceptibility along $x$ in body frame of plate: & $\chi_x$ & $-85\times 10^{-6}$ & NA \\
Magnetic Susceptibility along $y$ in body frame of plate: & $\chi_y$ & $-85\times 10^{-6}$ & NA \\
Magnetic Susceptibility along $z$ in body frame of plate: & $\chi_z$ & $-450\times 10^{-6}$ & NA \\\hline
Electrical Conductivity along x in body frame of plate: & $\sigma_x$ & 200 & ${\rm S/m}$ \\
Electrical Conductivity along y in body frame of plate: & $\sigma_y$ & 200 & ${\rm S/m}$ \\
Electrical Conductivity along z in body frame of plate: & $\sigma_z$ & 200 000 & ${\rm S/m}$ \\\hline
Checkerboard Cube Magnet Sidelength: & $magnetlength$ & 12.7 & $\rm mm$ \\
Checkerboard Magnet Magnetization: & $M_0$ & $1.18\times 10^6$ & ${\rm A/m}$ \\
Sidelength of square graphite solid slab: & $platedims_{xy}$ & $10$ & ${\rm mm}$ \\
Thickness of graphite solid slab:\tnote{a} & $platedims_z$ & $0.77$ & ${\rm mm}$ \\
{\bf Equilibrium} height of solid plate:\tnote{b} & {\bf $\bf z_{equil}$ } & 1.46 & ${\rm mm}$ \\
Density of Graphite: & $\rho_{graphite}$ & $2070$ & $\rm kg/m^3$ \\
Mass of Graphite resonator:
& $m$ & $79-147$ & $\rm mg$ \\
& &  &      \\ \hline\hline
\end{tabular}
\begin{tablenotes}
\item[a]The thickness change from area to area due the coarse surface of graphite plates.
\item[b]From top surface of the magnets to the center of the plate. It also changes with different resonators.
\end{tablenotes}
\end{threeparttable}
\end{center}
\vspace{-7mm}

\end{table*}

%